\documentclass[preprint,12pt]{elsarticle}
\usepackage{latexsym, graphicx, epsfig, amsmath, amssymb, amsfonts}
\usepackage{amsmath}
\usepackage{amssymb}
\usepackage{graphicx}
\usepackage{subfig}
\usepackage{tabu}
\usepackage{algorithm}
\usepackage{algpseudocode}
\usepackage{adjustbox}
\usepackage{bbold}
\usepackage{bm}
\usepackage{textcomp}
\usepackage[thinc]{esdiff}
\usepackage[dvipsnames]{xcolor}
\usepackage{comment}

\DeclareMathOperator*{\argmin}{argmin}

\begin{document}
\begin{frontmatter}

\title{Simulating progressive intramural damage leading to aortic dissection using an operator-regression neural network}
\author{Minglang Yin\textsuperscript{ab}}
\author{Ehsan Ban\textsuperscript{c}}
\author{Bruno V. Rego\textsuperscript{c}}
\author{Enrui Zhang\textsuperscript{d}}
\author{\\Cristina Cavinato\textsuperscript{c}}
\author{Jay D. Humphrey\textsuperscript{c}}
\author{George Em Karniadakis\textsuperscript{bd}\corref{cor1}}
\cortext[cor1]{Corresponding author: george\_karniadakis@brown.edu}
\address[a]{Center for Biomedical Engineering, Brown University, Providence, RI 02912}
\address[b]{School of Engineering, Brown University, Providence, RI 02912}
\address[c]{Department of Biomedical Engineering, Yale University, New Haven, CT  06520}
\address[d]{Division of Applied Mathematics, Brown University, Providence, RI 02912}

\begin{abstract}
Aortic dissection progresses via delamination of the medial layer of the wall. Notwithstanding the complexity of this process, insight has been gleaned by studying \textit{in vitro} and \textit{in silico} the progression of dissection driven by quasi-static pressurization of the intramural space by fluid injection, which demonstrates that the differential propensity of dissection can be affected by spatial distributions of structurally significant interlamellar struts that connect adjacent elastic lamellae. In particular, diverse histological microstructures may lead to differential mechanical behavior during dissection, including the pressure--volume relationship of the injected fluid and the displacement field between adjacent lamellae. In this study, we develop a data-driven surrogate model for the delamination process for differential strut distributions using DeepONet, a new operator--regression neural network. The surrogate model is trained to predict the pressure--volume curve of the injected fluid and the damage progression field of the wall given a spatial distribution of struts, with \textit{in silico} data generated with a phase-field finite element model. The results show that DeepONet can provide accurate predictions for diverse strut distributions, indicating that this composite branch-trunk neural network can effectively extract the underlying functional relationship between distinctive microstructures and their mechanical properties. More broadly, DeepONet can facilitate surrogate model-based analyses to quantify biological variability, improve inverse design, and predict mechanical properties based on multi-modality experimental data.
\end{abstract}

\end{frontmatter}
\textbf{keywords:} Aortic Dissection, Damage Mechanics, Operator Regression, Neural Networks, Phase Field Finite Elements, Soft Tissue
\newpage

\section{Introduction}
\label{sec:introduction}
\begin{figure}
    \centering
	\includegraphics[width=0.8\textwidth]{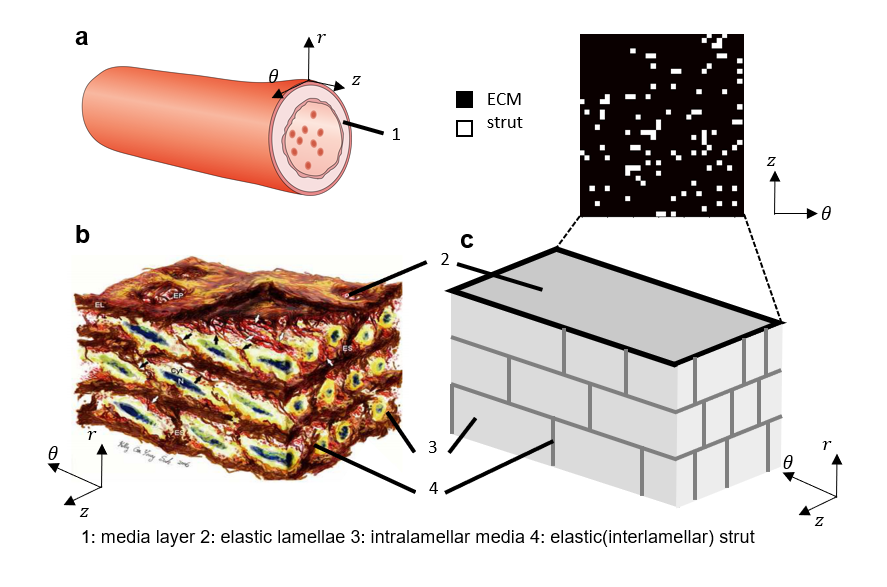} 
    \caption{
    \textbf{Sketch of the tissue microstructure.} (a) The aortic wall is composed of three layers: intima, media, and adventitia. (b) Artistic rendering of aortic medial microstructure by O'Connell et al.~\cite{o2008three}, reproduced with permission, and (c) its representative volume element for the finite element simulation. A top view of the intralamellar elastic struts (in white) is provided. 1: media layer, 2: elastic lamellae, 3: intralamellar media, 4: elastic (interlamellar) strut. $z$: axial, $\theta$: circumferential, $r$: radial. }
    \label{fig:elastic_unit}
\end{figure}

Aortic dissection, a cardiovascular condition resulting in high mortality, manifests within the medial layer of the aortic wall due to physical separation of the lamellar units (Fig.~\ref{fig:elastic_unit}(a)). The normal medial microstructure in Fig.~\ref{fig:elastic_unit}(b) reveals a ``sandwich'' structure wherein the elastic lamellae delimit intralamellar media (cells and extracellular matrix) and are connected by interlamellar struts (fibrillins, elastin, and collagens)~\cite{o2008three}. One possible outcome for an artery undergoing dissection is that the tear turns inward and forms a false lumen, while another possibility involves the tear turning outward and causing vessel rupture; the latter scenario can be lethal. Although most studies have focused on the stage with a developed false lumen, there has been little investigation of the initiation and propagation of aortic dissection from a mechanistic perspective. This knowledge gap prevents further predictive modeling endeavors.


There exist several hypotheses aiming at bridging this knowledge gap. One hypothesis states that dissection may arise from an intimal defect; that is, a defect within the luminal surface. Following this line, many insightful \textit{in silico} works assessed the risk of a false lumen for further dissection~\cite{peng2019patient} and investigated the stress distribution for an artery with flaps~\cite{rajagopal2007towards,karmonik2013computational, polanczyk2018computational, karmonik2012longitudinal, nathan2011pathogenesis, gao2006fluid, martin2015patient}. These works provided new insights into the propagation of a pre-existing defect, but did not address the initiation and propagation of aortic dissection. Another theory hypothesizes that a dissection may occur at or near a region with a concentrated stress field; for example, the subclavian branch of the aorta has higher stresses due to the sharp geometric and material discontinuity~\cite{fitzgibbon2020numerical}.

Additionally, it has been hypothesized and demonstrated that aortic dissection may arise from focal accumulations of glycosaminoglycans (GAGs)~\cite{humphrey2013possible, roccabianca2014biomechanical}, which are highly negatively charged and thus imbibe and sequester water. The combination of Gibbs-Donnan swelling pressure and internal stresses on the arterial wall can predispose to dissection, leading to a higher probability of rupture. Based on this line, a series of \textit{in silico} studies have used smoothed particle hydrodynamics~\cite{rausch2017modeling, ahmadzadeh2019modeling, ahmadzadeh2018particle} and standard finite element modeling~\cite{roccabianca2014biomechanical, roccabianca2014computational} to illustrate the role of accumulations of GAGs on the progression of dissection. Seminal work by Roach and colleagues focused instead on directly measuring the intramural (blood) pressure needed to propagate an initial medial defect within the thoracic aorta~\cite{he1994composition, maclean1999role, carson1990strength, tam1998effect, van1987factors, roach1994variations}, which provided further insights into dissection from \textit{in-vitro} experiments. They consistently found that static pressures needed to be 500~mmHg or more to dissect the normal wall and examined the mechanical properties of the porcine aorta at different locations by injecting fluid quasi-statically. Their works have shed light on the mechanism and characteristics of GAG-induced dissection initiation and propagation.


Most recently, Ban et al.~\cite{ban2021differential} developed a phase-field model to investigate aortic dissection with a histologically motivated microstructure; specifically, they simulated studies by Roach and colleagues~\cite{roach1994variations} focusing on the pressure-volume (P-V) curve associated with the intramural fluid that initiates and drives intramural damage. The realistic microstructure (Fig.~\ref{fig:elastic_unit}(b)) is simplified and represented in the phase-field model; Fig.~\ref{fig:elastic_unit}(c) shows the top view of an elastic strut distribution, which varies substantially from sample to sample. The variability in the elastin architecture leads to a correspondingly large difference in mechanical properties, dissection propensity, and severity from sample to sample and along the aorta (which is highest in the ascending thoracic aorta and lower in the abdominal aorta). Yu et al.~\cite{yu2020avalanches} further showed numerically and experimentally that the pressure drops induced by interlamellar damage to collagen follow a power-law behavior, indicating a possibility of predicting dissection based on the microstructure of the extracellular matrix. The power-law behavior was confirmed by Ban et al~\cite{ban2021differential}. Conversely, Holzapfel and colleagues used a phase-field modeling approach to investigate the propagation of dissection in mode I tearing~\cite{tong2016mechanical, gasser2006hyperelastic, sommer2008dissection}. The computational cost associated with performing enough finite element simulations to empirically uncover precise relationships between tissue microstructure and dissection propagation is prohibitive, however. Hence, there is a critical need to develop a reliable surrogate model of the process by which dissection propagates within the aortic wall. Such a model will greatly facilitate further \textit{in silico} investigations into this pathology, with the ultimate goal of developing computational tools to clinically assess and predict dissection propagation on a patient-specific basis, while also quantifying uncertainties associated with the resulting prognoses.

The potential of machine learning has grown rapidly in recent years, and the development of machine learning-based surrogate models for fast and accurate assessments has emerged as a key application of interest~\cite{brunton2020machine,lee2020model,qian2020lift}. Importantly, physics-informed neural networks (PINNs)~\cite{lagaris1998artificial, raissi2019physics} serve as one of the most promising models in scientific machine learning. PINNs penalize the residual of governing equations for the system in the loss function, where the partial derivatives are computed through automatic differentiation. This framework has been applied to predict flow fields~\cite{jin2021nsfnets, mao2020physics} and their variants~\cite{jagtap2020conservative}, fracture progression~\cite{goswami2020transfer}, and material property inference~\cite{yin2021non, zhang2020physics} among many other applications~\cite{cai2021physics,Karniadakis2021}. In biomedical engineering, PINNs have shown potential in cardiac activation mapping~\cite{sahli2020physics}, inferring arterial boundary conditions~\cite{kissas2020machine}, and inferring thrombus material properties~\cite{yin2021non} as summarized in~\cite{alber2019integrating}. Yet, for a system with different boundary/initial conditions, one has to retrain the network for each case, making the algorithm time-consuming. Hence, there is a pressing need to develop models that can learn the operator level of mapping between functions; that is, predicting the physical system under diverse boundary/initial conditions. Lu et al.~\cite{lu2021learning} developed the data-driven framework ``DeepONet,'' an operator--regression neural network that learns the mapping between functions with theoretical guarantees. DeepONet has shown its effectiveness in predicting multiscale and multiphysics problems \cite{cai2020deepm}, bubble growth~\cite{lin2020operator}, and crack prediction in brittle materials~\cite{goswami2021physics}. Li et al.~\cite{li2020fourier} developed a Fourier neural operator for learning parametric partial differential equations. 

In this work, we consider the phase-field intralamellar damage model developed in~\cite{ban2021differential} as the prototypical model to investigate dissection for a heterogeneous arterial wall. The mechanical process in the fluid-injection initiated arterial damage problem is well-characterized by its underlying P-V curve and damage progression. To the best of our knowledge, this is the first attempt to predict dissection progression and mechanical behavior in a heterogeneous aortic wall using scientific machine learning. We will demonstrate that details of the characteristic pressure drops can be well-characterized by incorporating the current damage field; otherwise, the model prediction will lead to a mean-field average without detailed pressure drops. Moreover, we investigate the model performance on predicting the damage field given the observable displacement field. 

The paper is organized as follows. In section~\ref{sec:methods}, we introduce the details of the phase-field model that generates synthetic data and details of DeepONet. In section~\ref{sec:results}, we introduce the data preparation, followed by the prediction results of DeepONet for the corresponding P-V curve and the damage field. Moreover, we evaluate the effect of network structure on testing errors. In section~\ref{sec:discussion}, we conclude by interpreting the results as well as discussing further applications of the presented model. 

\section{Methods}
\label{sec:methods}
\subsection{Phase Field Model}
Herein, we employ a validated phase-field finite-element method~\cite{gultekin2019computational} to describe progressive damage in the native aortic wall. Unlike previous particle methods~\cite{rausch2017modeling, ahmadzadeh2018particle}, the phase-field model gives a continuum description of a discontinuous tear. The phase field $\phi(\textbf{x}, t) \in [0, 1]$ satisfies
\begin{equation}
    \phi(\textbf{x}, t) = 
    \begin{cases}
        1, & \quad \textnormal{intact region,} \\
        0, & \quad \textnormal{torn region}.
    \end{cases}
\end{equation}
We adopt the same phase-field model developed in ~\cite{ban2021differential} to describe dissection progression within the heterogeneous aortic media driven by quasi-static injection of fluid. In modeling the delamination, we sought to find the displacement $\textbf{\textit{u}}$, pressure $p$, Lagrange multiplier $m$, and phase field $\phi$ by the minimization of total energy $E_{\text{total}}$, 
\begin{equation}
    (\textbf{\textit{u}}_{i}, p_{i}, m_{i}, \phi_{i}) = \argmin_{\textbf{\textit{u}},p,m,\phi} E_{\text{total}}(\textbf{\textit{u}},p,m,\phi),
\end{equation}
at each increment of injection step, $i$. The total energy $E_{\text{total}}$ is expressed as
\begin{equation}
    E_{\text{total}} = E_{\text{deform}} + E_{\text{tear}} + E_{\text{pressurized-fluid}} ,
\end{equation}
where these three terms represent the energy contributions from elastic deformation, tearing, and injection. 

The native aortic wall is modeled as a hyperelastic material, with the strain energy density function $W_{\text{wall}}$ used to calculate effects of elastic wall displacement. Specifically, $W_{\text{wall}}$ is modeled as a Fung exponential stiffening material~\cite{holzapfel2000new,roccabianca2014computational}:
\begin{equation}\label{Fung-type}
    W_{\text{wall}} = \frac{c}{2}(\lambda_{1}^{2} + \lambda_{2}^{2} + \lambda_{3}^{2} - 3) + \sum_{i=1}^{4}\frac{c_{1}^{i}}{4c_{2}^{i}}(e^{c_{2}^{i}[ (\lambda^{i})^{2} - 1 ]^2} - 1),
\end{equation}
where $c$, $c_{1}^{i}$, and $c_{2}^{i}$ are material constants. Here, $\lambda_{1}$, $\lambda_{2}$, and $\lambda_{3}$ are principal stretch ratios and $\lambda^{i}$ are stretch ratios in one of the four primary directions associated with the intramural fibrous constituents, mainly collagens ($i = 1, 2, 3$, and $4$ are axial, circumferential, and two diagonal directions, respectively). We use material properties reported in Ref.~\cite{roccabianca2014quantification} based on previous biaxial experiments on human aortas \cite{geest2004age}. The stiffness of the struts and the matrix differ by a factor of 20 such that their arithmetic mean yields the properties of the homogeneous wall.

A similar approach to the perturbed Lagrangian method is applied to constrain the nearly incompressible condition of the wall material:
\begin{equation}
    W_{\text{vol}} = -(1-\phi)^{3}p(\lambda_{1}\lambda_{2}\lambda_{3} - 1) + \frac{p^2}{2\epsilon},
\end{equation}
where ${p^2}/{(2\epsilon)}$ serves as a regularization term, with $\epsilon \gg 1$. Hence, the total energy of deformation of the wall is:
\begin{align}
    E_{\text{deform}} &= \int_{V} [( (1-\phi)^{2} + \varepsilon)W_{\text{wall}} + W_{\text{vol}}] \, dV, 
\end{align}
where the term $(1-\phi)^2$ serves as the degrading function for the damage region, and $\varepsilon \ll 1$ serves as a regularization term.

The tearing energy can be expressed as: 
\begin{equation}
    E_{\text{tear}} = \frac{3G_{c}}{8}\int_{V} \left( \frac{\phi^{2}}{l} + l\left|\nabla\phi\right| ^{2} \right) \, dV,
\end{equation}
with the critical energy release rate of tearing $G_{c}$ set respectively, as $5.77$ or $0.20$ $J/m^{2}$ for the radial struts and surrounding matrix to reproduce the pressure--volume curve in~\cite{roach1994variations}. The dissection advances by the injection of pressurized fluid. In this work, we constrain over the total volume of the injected fluid by using a Lagrange multiplier $m$ to prescribe the total volume of fluid, $V_{\text{injection}}$, with
\begin{equation}
    E_{\text{pressurized-fluid}} = m \left( \int_{V}\phi\lambda_{1}\lambda_{2}\lambda_{3} \, dV - V_{\text{injection}} \right).
\end{equation}
Numerical viscous-like damping was used to facilitate the numerical solution.

We iteratively minimize the total energy at each loading step by taking the variational derivative in the direction of $\textbf{\textit{u}}$, $p$, $m$, and $\phi$. We use the Bubnov--Galerkin method and the FEniCS~\cite{alnaes2015fenics} finite element package to solve the resulting nonlinear weak formulation.


\subsection{Histologically Motivated Microstructure}

\begin{figure}
    \centering
	\includegraphics[width=0.8\textwidth]{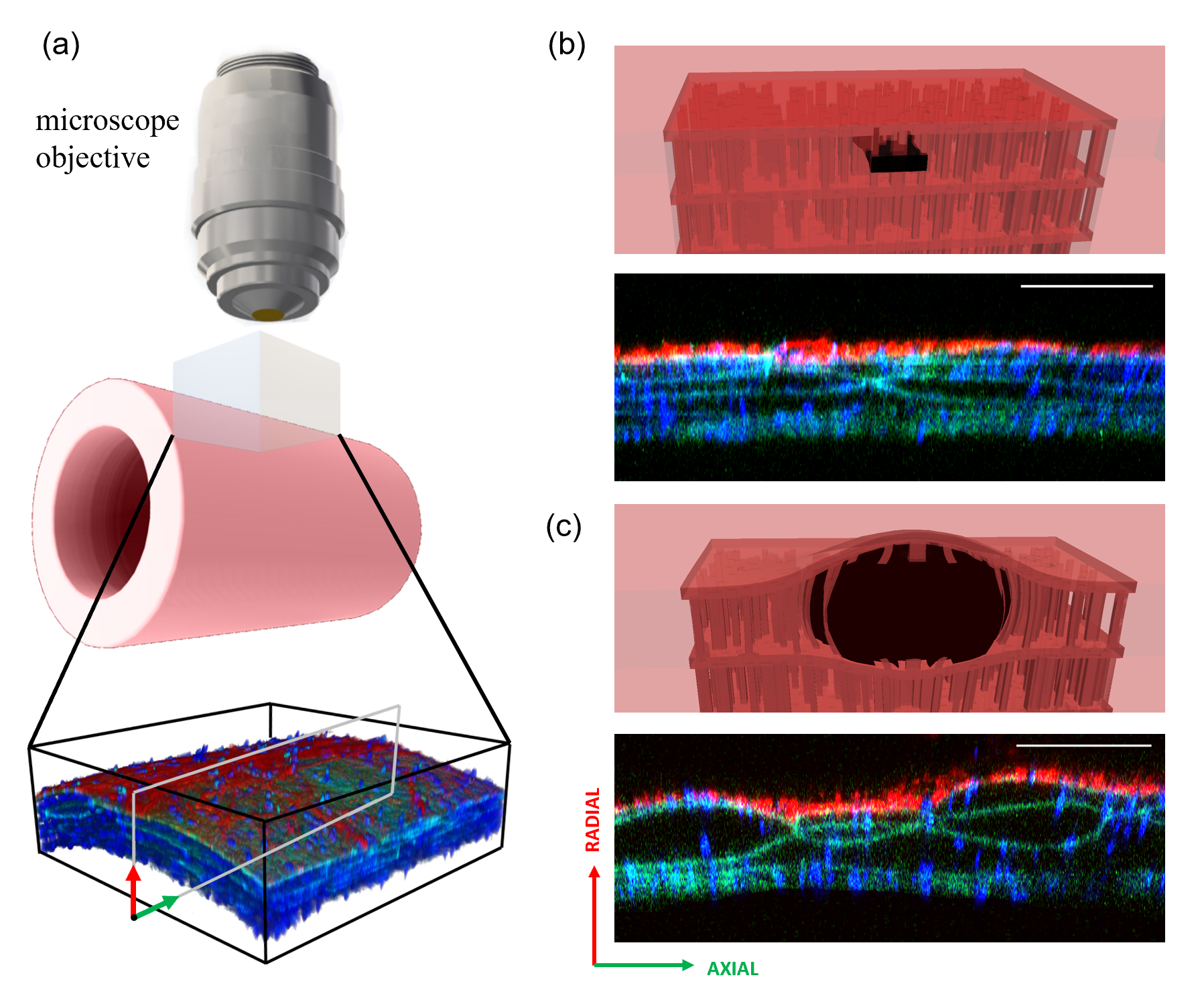}
    \caption{\textbf{Histologically motivated microstructure of the murine aortic wall under biaxial loading.} (a) 3D imaging of the murine arterial wall by multiphoton microscopy. Aortic wall (b) initiating and (c) propagating intralamellar delaminations, including histologically motivated microstructure for the phase-field model. Green: elastic lamellae. Blue: Smooth muscle cell nuclei. Red: Collagen in the adventitia. Scale bar: 100 µm.}
    \label{fig:multiphoton}
\end{figure}

Microstructures generated for the phase-field model are motivated by the realistic histology of the murine aortic wall. We present a 3D image of the wall acquired by multiphoton microscopy \cite{cavinato2021evolving} to illustrate the ``sandwich'' structure wherein the medial layer is composed, on average, of 5-6 layers of elastic lamellae indicated by the green pixels in Fig.~\ref{fig:multiphoton}. In Fig.\ref{fig:multiphoton}(a), we show a subvolume of reconstructed murine aorta under \textit{ex vivo} biaxial loading equivalent to physiological conditions. We compare the cross-section of an artificial-generated microstructure used in the phase-field simulations and a real aortic wall in (b), showing a qualitative consistency between the two geometries. Different components of the arterial wall are distinguished by colors: adventitial collagen is denoted by red, elastic lamellae by green, and smooth muscle cell nuclei by blue.

Admittedly, we notice that the real geometry is not homogeneous in the sense that the elastic lamellae in the phase-field geometry can be clearly identified. Nonetheless, it is still representative in that the strut, matrix, and lamellae structure are preserved. In Fig.~\ref{fig:multiphoton}(c), we compare an undergoing aortic dissection between the phase-field model and real arterial wall, where a visible false lumen can be observed from both images.

\subsection{DeepONet}

\begin{figure}
    \centering
	\includegraphics[width=0.9\textwidth]{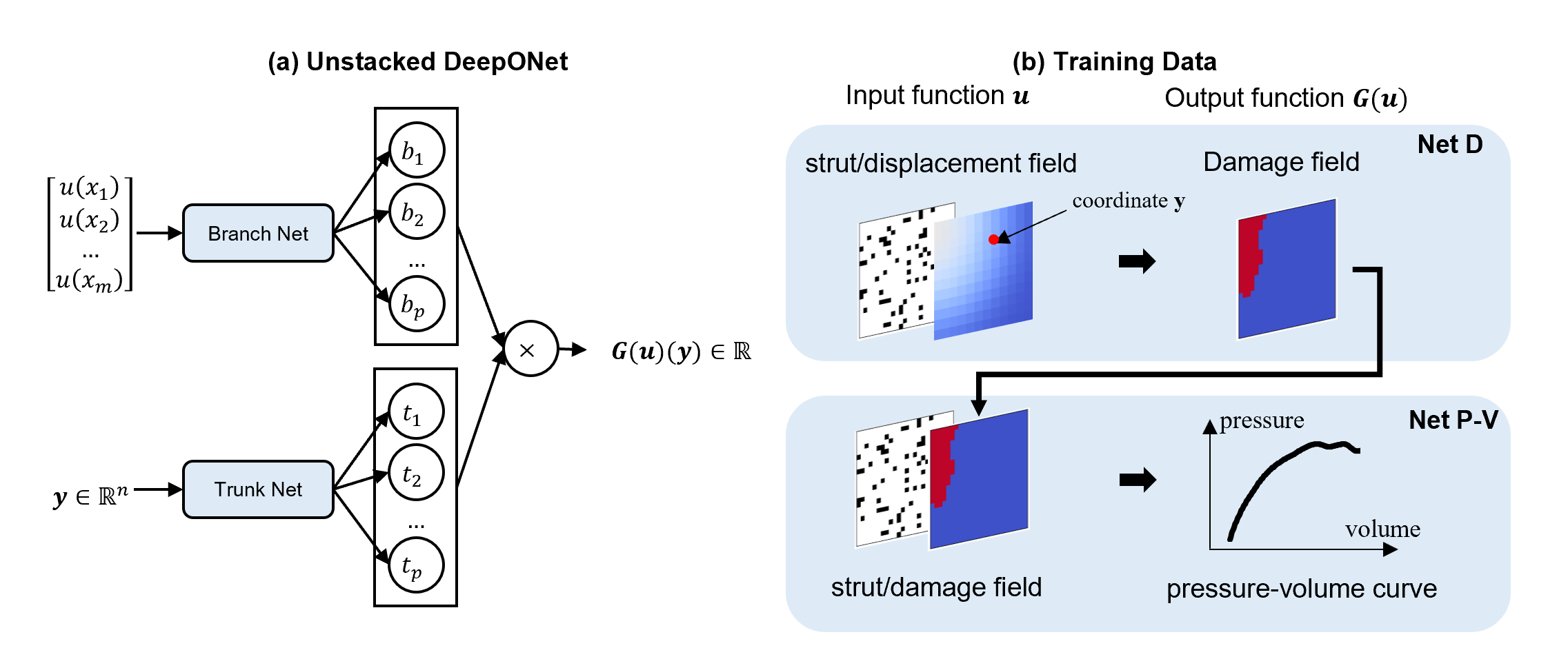} 
    \caption{\textbf{A schematic of DeepONet for fluid injection-induced dissection problems and its training data.} (a) An unstacked DeepONet for learning an operator $G:u\rightarrow G(u)$ takes two inputs $[u(x_{1}), u(x_{2}),...,u(x_{m})]$ and $y$ and returns $G(u)(y)$, which is a scalar. The trunk net is a fully-connected neural network (FNN) whereas the architecture of the branch net varies depending on the problem. (b) Training data for the problem. First, we train \textit{Net D} whose input $u$ contains the elastic struts map and the current surface displacement map to predict the damage field at the current state. Next, the predicted damage field is taken as an input for \textit{Net P-V} for predicting the corresponding pressure-volume curve for the injection fluid. Each network is trained separately and coupled for prediction.}
    \label{fig:deeponet}
\end{figure}

In this section, we summarize the general architecture of DeepONet; we refer the reader to \cite{lu2021learning} for more details. DeepONet was proposed for learning nonlinear operators, by mapping input functions into corresponding output functions. Let $G$ be a nonlinear operator taking an input function $u$ and yielding the output function $G(u)$. Evaluation of the function $G(u)$ at $y\in\mathbb{R}^{n}$ is a real number and can be denoted as $G(u)(y)$, where $y$ typically is vector containing coordinates and other information. In practice, the input function $u$ is represented discretely by its value $[u(x_1), u(x_2),...,u(x_m)]^\text{T}$ at a finite set of locations $\{x_i\}_{i=1}^m$. The architecture of the (unstacked) DeepONet is shown in Fig. \ref{fig:deeponet}(a). The branch net and the trunk net take the function $u$ (in the form of $[u(x_1), u(x_2),...,u(x_m)]^\text{T}$) and the vector $y$ as inputs, and yield intermediate outputs $[b_1,b_2,...,b_p]^\text{T}$ and $[t_1,t_2,...,t_p]^\text{T}$, respectively. In the last layer, DeepONet merges these two outputs by taking their pointwise multiplication:
\begin{equation}
    G(u)(y) \approx \sum_{k=1}^{p}b_{k}t_{k} + b_{0},
\end{equation}
where we include an additional bias term $b_0\in\mathbb{R}$ as a trainable variable as it may reduce the generalization error. With such a setup, the $i$-th sample from the training/testing dataset gives a triplet $[u^{(i)}, y, G(u^{(i)})]$ (branch input, trunk input, and output) with the following structure 
\begin{gather}
    \begin{bmatrix}
        \begin{bmatrix}
        u^{(i)}(x_{1}) & u^{(i)}(x_{2}) & \cdots & u^{(i)}(x_{m}) \\
        u^{(i)}(x_{1}) & u^{(i)}(x_{2}) & \cdots & u^{(i)}(x_{m}) \\
        \vdots & \vdots & \ddots & \vdots \\
        u^{(i)}(x_{1}) & u^{(i)}(x_{2}) & \cdots & u^{(i)}(x_{m}) \\
        \end{bmatrix}, &
        \begin{bmatrix}
        y_{1} \\
        y_{2} \\
        \vdots \\
        y_{N} \\
        \end{bmatrix}, &
        \begin{bmatrix}
        G(u^{(i)})(y_{1}) \\
        G(u^{(i)})(y_{2}) \\
        \vdots \\
        G(u^{(i)})(y_{N}) \\
        \end{bmatrix}
   \end{bmatrix},
\end{gather}
where evaluation of the function $u^{(i)}$ at $[x_{1}, x_{2}, \ldots,x_{m}]$ is copied $N$ times to match the evaluation of $G(u)$ at points $y_{j}\in \mathbb{R}^{n}$ ($j = 1,2,\ldots,N$). The dimension of each term on the right hand side is $(N, m)$, $(N, n)$, and $(N, 1)$.

We develop two DeepONets (\textit{Net D} and \textit{Net P-V} in Fig.~\ref{fig:deeponet}(b)) for two different tasks: predicting damage progression and corresponding pressure--volume curves. The architecture of the branch net and the trunk net is selected to adapt the structure of their input data. In this study, the inputs of the branch net ($[u(x_1), u(x_2),...,u(x_m)]^\text{T}$) are a stack of 2D images indicating in-plane fields of physical quantities, including the strut distributions, damage progression, and displacement profiles depending on specific tasks. In view of this, we mainly adopt a convolutional neural network (CNN), an efficient architecture for processing images, as the architecture of the branch net. A fully-connected neural network (FNN) is also considered for the branch net for comparison. For the trunk net, we always choose the FNN model due to small dimensions (no more than two) of the input $y$, which can be the in-plane spatial coordinates or the injected fluid volume depending on specific tasks. We will provide more details regarding the definition of the tasks in this study in section \ref{sec:results}.

We adopt the mean squared error (MSE) to measure the discrepancy between the model prediction and the data. To avoid overfitting and to improve the generalization capability of the model, we introduce a $L_2$ regularization term in the loss function, which penalizes large values of weights and biases in the trunk net and the final output layer. Therefore, the total loss $\mathcal{L}$ can be expressed as
\begin{equation}
    \mathcal{L} = \frac{1}{MN}\sum^{M,N}_{i,j=1}(G_{\text{model}}(u^{(i)})(y_j) - G_{\text{data}}(u^{(i)})(y_j))^{2} + \alpha \sum^{K}_{i=1} \xi_i^2,
\end{equation}
where $M$ is the size of the dataset, $G_\text{model}$ is the operator learned by the DeepONet, $G_\text{data}$ is the target operator, $\xi_i$ ($i=1,2,\ldots,K$) is the $i$-th trainable parameter of the trunk net and the output layer, and $\alpha$ is the weight of regularization ($10^{-5}$ or $10^{-6}$ in this work).

\section{Results}
\label{sec:results}
\subsection{Data Preparation}

\begin{figure}
    \centering
	\includegraphics[width=0.8\textwidth]{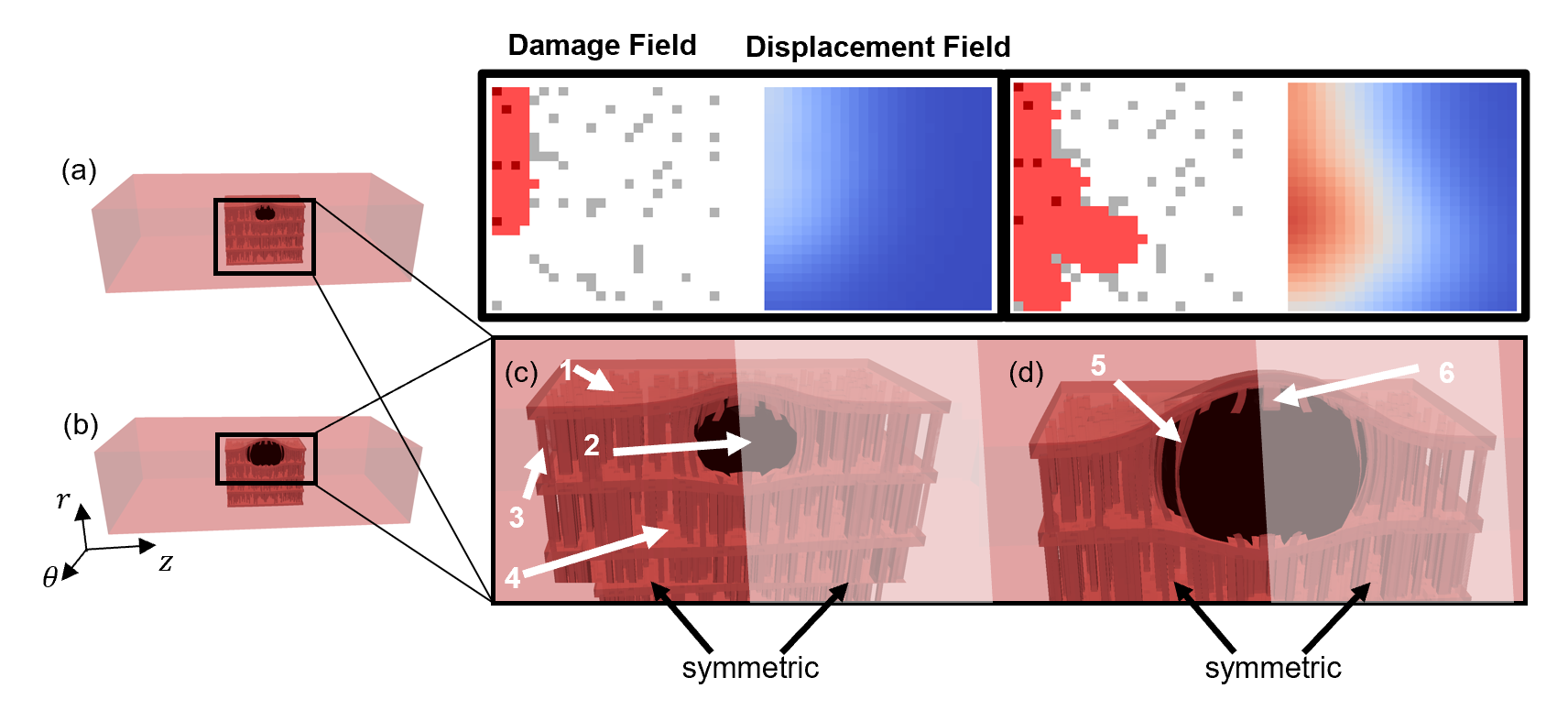} 
    \caption{\textbf{3D snapshots for injection-induced delamination in the phase-field model at (a) an early stage and (b) a progressed stage.} Note that (c, d) show enlarged views with details: (1) elastic lamellae, (2) damage region, (3) elastic struts, (4) interlamellar media, (5) a strut being torn, and (6) a broken strut. The computational geometry is symmetric with respect to the transparent part, hence only half of the damage and displacement fields are shown on the top for illustration.}
    \label{fig:dissection}
\end{figure}

\begin{figure}
    \centering
	\includegraphics[width=1.0\textwidth]{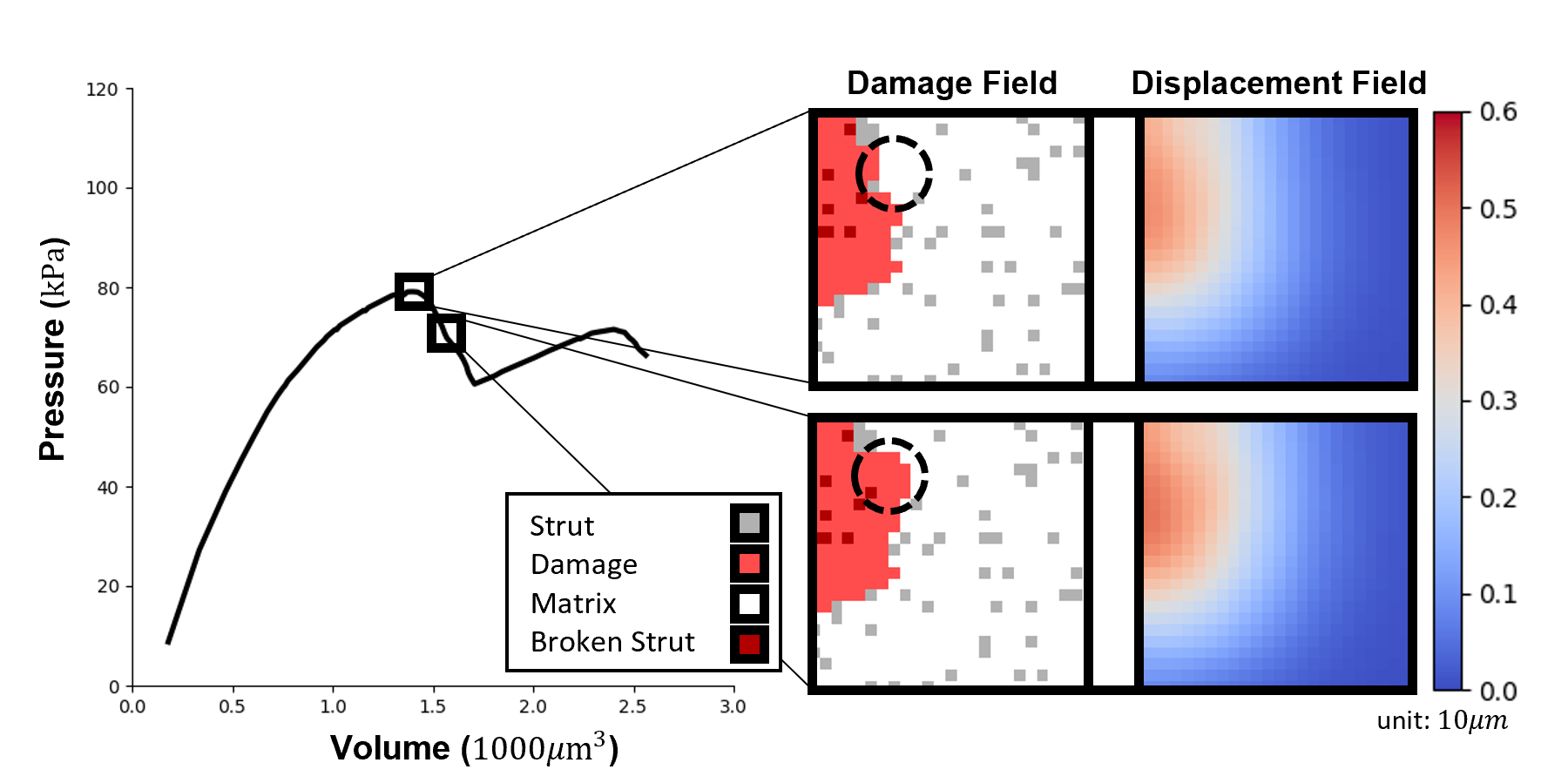} 
    \caption{\textbf{Pressure drop and its corresponding dissection state.} A pressure drop is induced by irreversible tearing of the elastic struts and matrix indicated in the circle, with their corresponding displacement fields on the right. The red region indicates the damaged area and the gray pixels are the elastic struts. The torn struts within the damage region are represented with darker red. The white pixels indicate intact matrix.}
    \label{fig:illustration}
\end{figure}

The phase-field finite element model provided synthetic training data for the DeepONet. This model consists of several elastic lamellae, modeling the medial layer (Fig.~\ref{fig:dissection}). Medial delamination is driven by pressurized fluid between two consecutive elastic sheets, with the fluid injected quasi-statically. Initially, increments of fluid volume elastically deform the wall, causing the top-most elastin sheet to bulge. At a critical pressure, the matrix begins to tear while the radial struts sustain the fluid pressure. The struts also resist fluid propagation both by keeping the adjacent sheets together and by acting as strong barriers. As the volume increases, however, a few of the struts snap, resulting in a sudden drop of pressure shown in Fig.~\ref{fig:illustration}. Given the experimentally observed stochastic distribution of the position of struts, this process results in a stochastic variation of pressure in response to the injection of intramural fluid ~\cite{ban2021differential}.

To generate the requisite training and testing datasets for DeepONet, we consider 2,100 phase-field solutions, each with a different stochastic distribution of strut position. In each case, the outputs from the finite element model include the P-V curve and two planar fields; the displacement field records the rising of the innermost medial elastic lamina and the damage field marks the tearing of tissue as the fluid progresses. Averaging over a section in the intralamellar space, where the fluid is injected, produces pixel-wise maps based on the finite element solution (Fig.~\ref{fig:dissection}). The injected volume in each case varies from $~1.3\times10^{3}\mu \textnormal{m}^{3}$ to $~3.5\times10^{3}\mu \textnormal{m}^{3}$.

\begin{figure}
    \centering
	\includegraphics[width=0.8\textwidth]{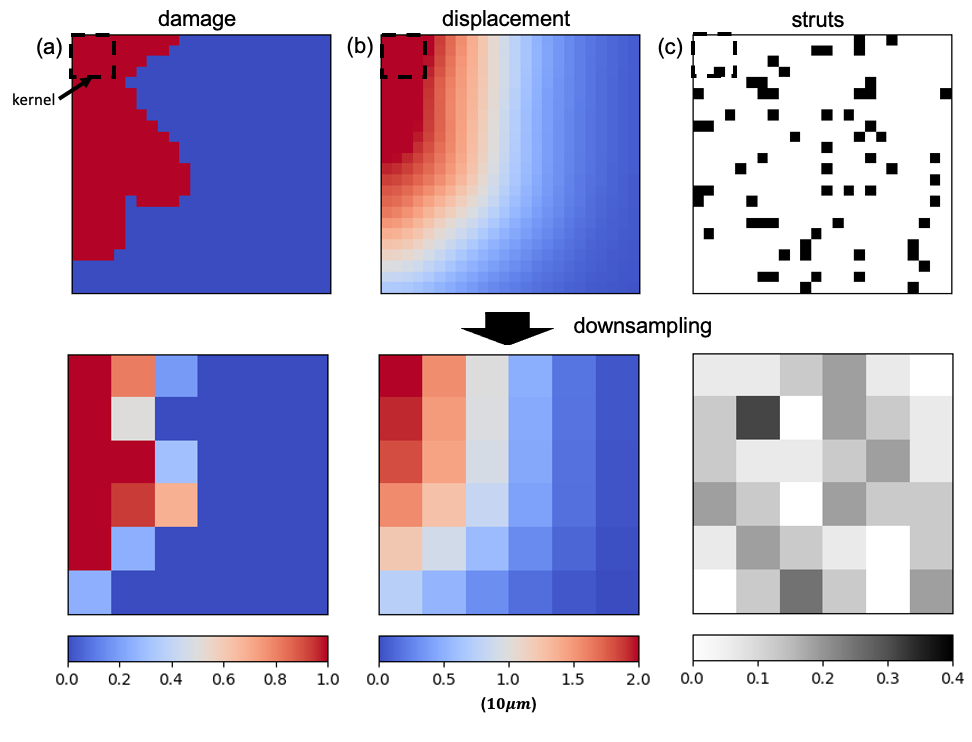} 
    \caption{\textbf{Image downsampling.} Damage, displacement, and strut images are downsampled by taking the average within a 4$\times$4 kernel. Note that the physical meaning for each field after downsampling is: (a) the percentage of damage within a pixel, (b) mean displacement, and (c) percentage of strut within a pixel.}
    \label{fig:downsampling}
\end{figure}

The maps, representing the damage fields, displacement fields, and strut distributions, are represented by \textbf{$36\times36$} 2D images. In the damage field shown in Fig.~\ref{fig:downsampling}, the red pixels indicate the torn area with value 1, whereas the blue pixels represent the intact area with value 0. In Fig.~\ref{fig:downsampling}(c), which represents the heterogeneous microstructure of the arterial wall, the black pixels are struts with value 1 and the white pixels are surrounding matrix with value 0. The displacement field is continuous, however, where the red pixels stand for larger vertical displacement. In particular, to reduce the computational cost and facilitate a faster training process, we reduce the size of the original images from $36\times36$ to $6\times6$ by taking the average pixel within a $4\times4$ kernel.

\subsection{Damage Field Prediction (\textit{Net D})}\label{sec:Net_D}

\begin{figure}
    \centering
	\includegraphics[width=1.0\textwidth]{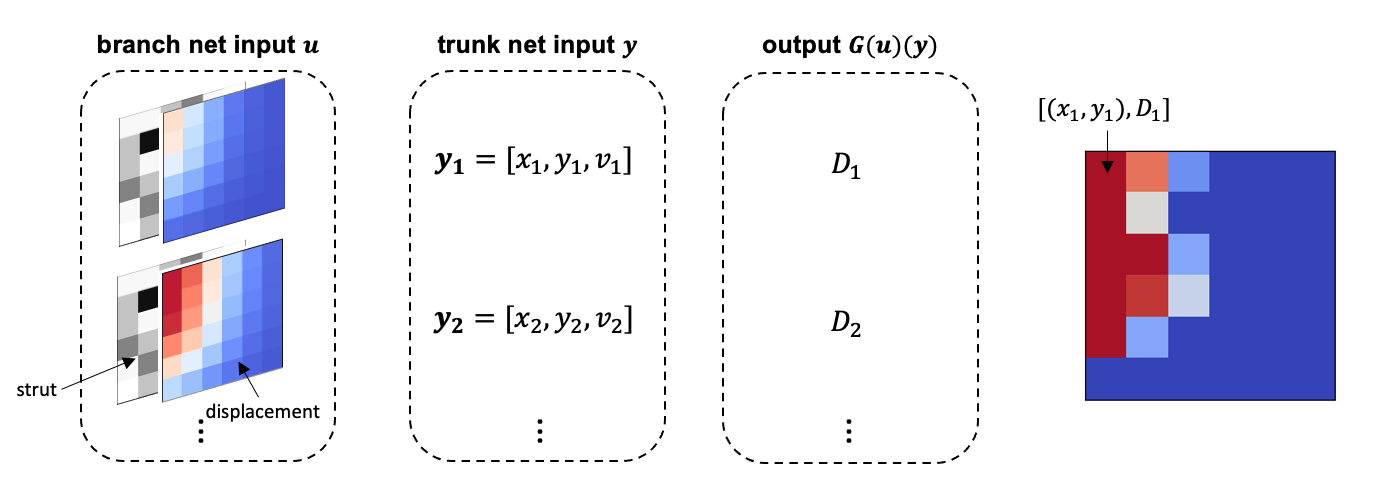} 
    \caption{\textbf{Triplets of a single training case for predicting damage progression.} A dataset triplet is composed of images $\textbf{u}$ (strut and displacement fields), a vector $\textbf{y}$ containing pixel coordinates $x_{i}, y_{j}$ and volume of injection $v_{i}$, and the damage percentage $D_{i}$.}
    \label{fig:damage_triplet}
\end{figure}

Although the displacement field can be experimentally observed \textit{in vitro}~\cite{bersi2020multimodality}, the damage field is a hidden quantity that is hard to observe directly. In this section, we infer the current damage field based on the displacement field using DeepONet. The triplet of the training dataset is presented in Fig. \ref{fig:damage_triplet} where the input and output are listed within. The trunk net input $y$ is a concatenation of pixel coordinates ($x_{i}, y_{j}$) and injection volume $v_{i}$ with the output $G(u)(y)$ the percentage of damage $D_{i}$ at that pixel. To evaluate model performance, we use 1,900 finite-element simulations as the training data and 200 more simulations for testing, where each case contains injection steps varying from 20 to 100. Each case contains a different microstructure with initial damage imposed at one boundary. As fluid is quasi-statically injected into the medial layer, the corresponding damage progression is distinctive due to the difference in the strut map. We list the model architecture in Table \ref{table:damage_network_size}.

\begin{table}
\centering
\begin{adjustbox}{width=0.7\textwidth}
 \begin{tabular}{c | c | c | c} 
 \hline
 Branch Net & Channels & Trunk Net & Width \\ \hline
 \hline
 CNN (small) & [4, 8, 12] & FNN (small) & [50, 50, 50]\\
 \hline
 CNN (large) & [10, 25, 50] & FNN (large) & [100, 100, 100]\\
 \hline
\end{tabular}
\end{adjustbox}
\caption{\textbf{Model architectures for predicting damage progression.} The trunk net is chosen as a fully-connected neural network (FNN) with three hidden layers while the branch net is a convolutional neural network (CNN) with three blocks. Each block contains a convolutional layer, a LeakyReLU layer, and a batch normalization layer.}
\label{table:damage_network_size}
\end{table}

\begin{figure}
    \centering
	\includegraphics[width=1.0\textwidth]{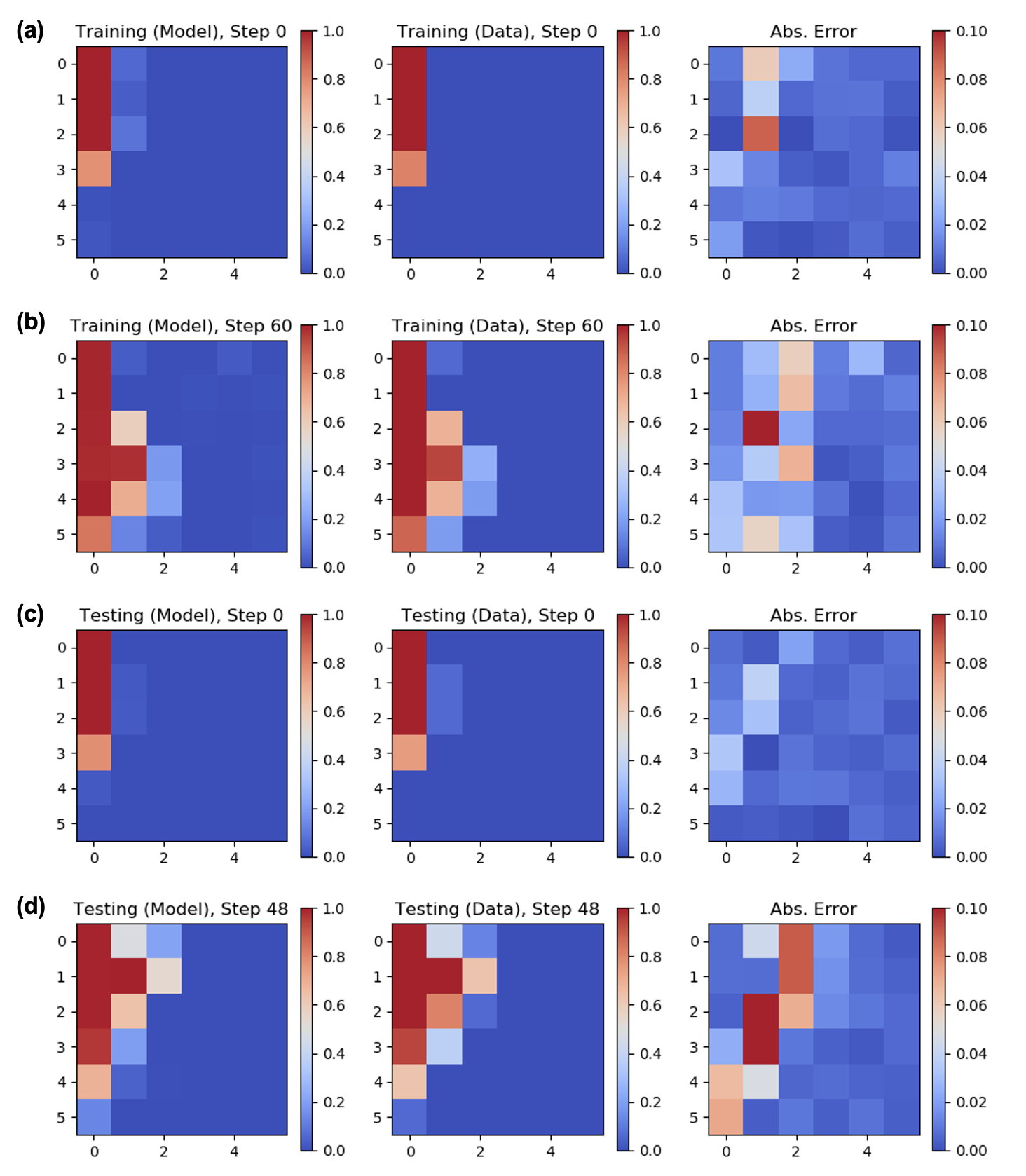} 
    \caption{\textbf{Training/Testing results of DeepONet for predicting damage progression.} Model predictions and true damage fields are shown in the first and second columns, while the third column presents their absolute difference. (a, b) show the results for a training case at steps 0 and 60, while (c, d) represent a testing case at steps 0 and 48. The damage field indicates the percentage of torn material in the region at a particular pixel.}
    \label{fig:damage}
\end{figure}

First, we qualitatively show the ability of DeepONet to predict the damage progression in Fig. \ref{fig:damage}(a); compare the model prediction and true damage data at the initial stage with the absolute error plotted in the third column. A later injection stage is plotted in Fig. \ref{fig:damage}(b). The largest absolute error for damage field prediction is below 10\% at the damage tip. In (c, d), we compare the inference results for the damage field from a testing case. We find an overall good agreement between model prediction and the true damage field at the initial step and last injection step (48). In general, the model predictions agree well with the simulations, with smaller errors at the initial loading step and larger errors at later steps.

\begin{figure}
    \centering
	\includegraphics[width=1.0\textwidth]{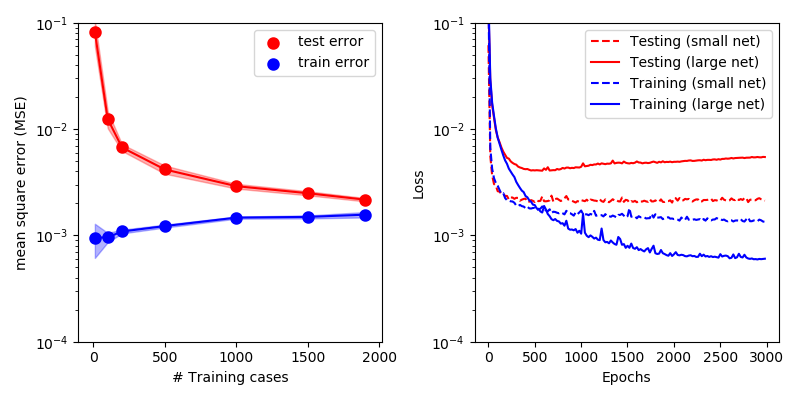} 
    \caption{\textbf{Errors for predicting damage fields against the number of training cases and epochs.} The network architecture is shown in Table \ref{table:damage_network_size}. The error bars show the one standard deviation from 5 runs with different initiation.}
    \label{fig:param_damage}
\end{figure}

Furthermore, we investigate the influence of the number of training cases on the prediction errors by training a network with a fixed architecture with different data. The number of training cases increases from 10 to 1,900, causing the testing error to drop from close to $O(10^{-1})$ to $O(10^{-3})$ with the training error increasing slightly as shown in Fig. \ref{fig:param_damage}(left). The values of testing error quickly converge to that of the training, indicating an improvement in performance with increasing training cases. Fig. \ref{fig:param_damage}(b) plots the errors against training epochs. We employ a large network structure with three hidden layers, each with 100 neurons, and a small network structure with three hidden layers and 50 neurons (Table \ref{table:damage_network_size}). By employing the small network, the testing error ends up at around 0.001 compared with 0.004 from the larger network, indicating the small network is better than the large network.

\subsection{Pressure--Volume Curve (\textit{Net P-V})} \label{sec:pv_curve}

\begin{figure}
    \centering
	\includegraphics[width=1.0\textwidth]{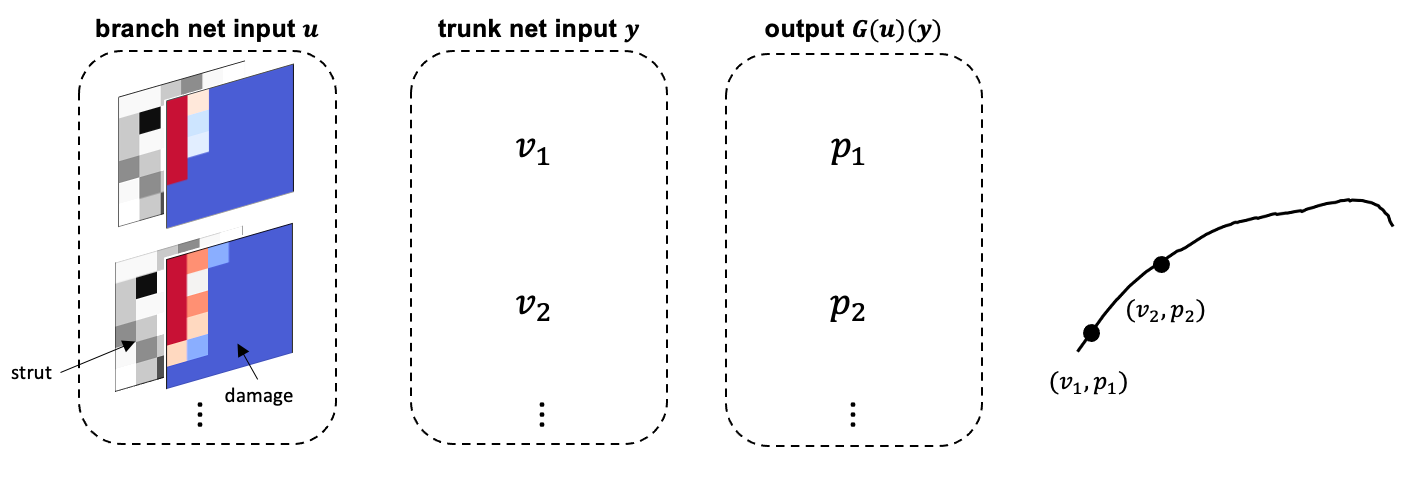} 
    \caption{\textbf{Triplets of a single training case for the P-V response.} A dataset triplet is composed of images $\textbf{u}$ (strut and damage fields), volume of injection $v_{i}$, and its corresponding pressure $p_{i}$. There are $n$ injection steps in a single case, corresponding to $n$ triplets.}
    \label{fig:pv_triplet}
\end{figure}

In this section, we investigate the performance of DeepONet in predicting P-V curves of injection-induced aortic dissection, which serves as the second part of the overall framework, predicting the corresponding P-V curves based on model predictions of \textit{Net D}. However, to train such a network, we use the true damage field from the data as training data, while incorporating the predicted damage field in our later experiments. The training dataset is a triplet with a structure listed in Fig. \ref{fig:pv_triplet}. It is important to highlight that the microstructure of a heterogeneous wall and current damage fields are represented by two $6\times6$ images as the input of the branch net. The trunk net input $y$ is the injected volume $v_{i}$ with the output $G(u)(y)$ for the corresponding fluid pressure $p_{i}$. We use the same training and testing data in \textit{Net D} to evaluate the model performance.

\begin{figure}
    \centering
	\includegraphics[width=0.8\textwidth]{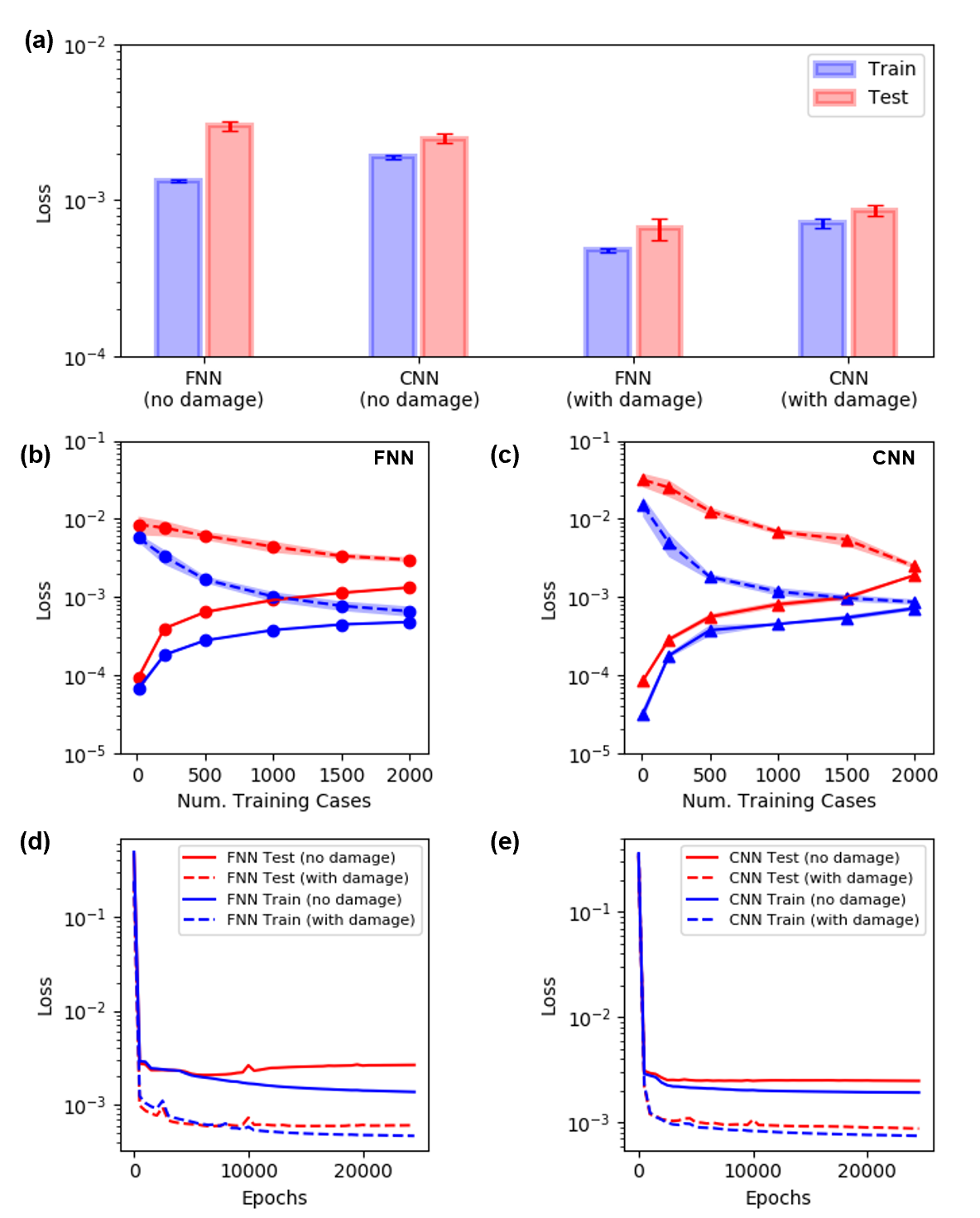} 
    \caption{\textbf{Learning the pressure--volume curve using different input data and trunk network architectures.} (a) Errors and deviations of different network architectures trained to learn the pressure--volume curve. The trunk net is chosen as a fully-connected network (FNN) for each run, while the branch net is chosen as a convolutional neural network (CNN) or a FNN, respectively. The networks trained with damage fields have a lower testing error than those without, given the same amount of training cases. The error bars show one standard deviation from 5 runs with different initiation. (b, c) The training and testing errors are plotted against the number of training cases. FNN and CNN (trunk net) trained with damage fields are plotted in blue, while those without damage fields are plotted in red. (d, e) Using the same color style, the losses vs training epochs are plotted for FNN and CNN, respectively. Note that the pressure values are normalized to $O(1)$ by dividing by 100.}
    \label{fig:arch_loss}
\end{figure}

First, we quantitatively investigate how the performance is influenced by the branch net architecture. We choose the branch net as a FNN or CNN trained with/without the damage field, respectively, with the detailed structure listed in Table \ref{table:pv_network_size}. We fix the trunk net with a structure $[1, 50, 50, 50]$. 
We plot the errors and deviations for the different network architectures for the P-V curve in Fig. \ref{fig:arch_loss}(a) and observe that errors associated with both CNN and FNN without the damage field are significantly larger than those with the damage field. In particular, the testing error is almost an order lower than for the case without damage data, at $O(10^{-3})$. Figs. \ref{fig:arch_loss}(b) and (c) show the training loss and testing loss versus the number of training cases. The testing error reduces from $O(10^{-2})$ to $O(10^{-4})$ with the number of training cases increasing from 10 to 1,900. The error bars show one standard deviation from 5 runs with different initiation. The results indicate that the DeepONet with FNN as the branch net has a slightly smaller training and testing error than that of CNN. Moreover, we plot the training and testing loss for FNN and CNN trained with/without information on the damage fields. The number of training and testing cases is again 1900 and 200, respectively. It is evident that both the training and testing errors with damage fields are smaller than the training/testing errors without the damage field. In addition, incorporating damage progression can avoid overfitting, which is pronounced in the testing error history (gray line) shown in (d) and (e).

\begin{figure}
    \centering
	\includegraphics[width=0.9\textwidth]{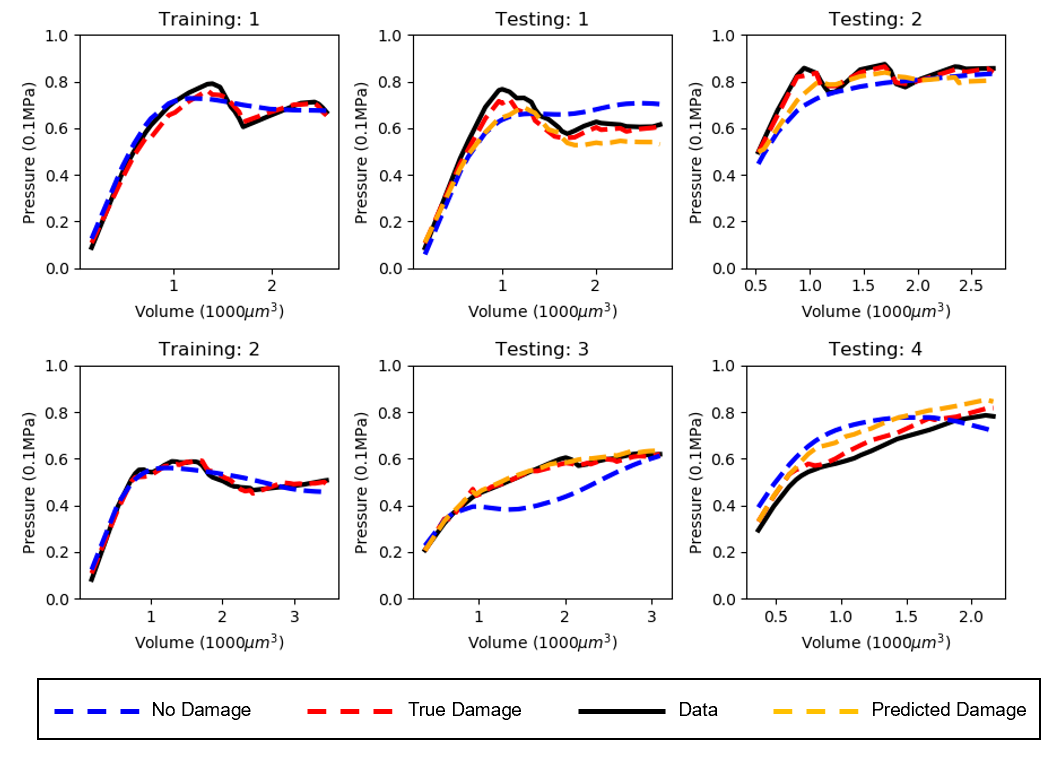}
    \caption{\textbf{DeepONet for predicting P-V curves.} Model predictions (dashed lines) for training cases 1 and 2 are compared with the true data (black line, first column). In particular, model predictions with and without damage fields are plotted in red and blue, respectively. The middle and right columns compare the testing data with models trained with(red)/without(blue) the current damage field. Notice that the orange line represents the predicted P-V curves based on the prediction of \textit{Net D}. The branch net is a FNN trained with 1900 training cases. The true data were generated from the phase-field finite element solver.}
    \label{fig:pv_curve}
\end{figure}

Then, we compare the model prediction and the phase-field model results (black lines) in Fig. \ref{fig:pv_curve}. The branch net is a FNN with structure [64, 50, 50, 50] while the trunk net is a FNN with [1, 50, 50, 50]. We compare the model predictions from DeepONet models trained with different types of data: one with the true/predicted damage field and the other one without. It is not practical to have the true damage field as input since it is a hidden quantity, we thereby employ the predicted damage fields from \textit{Net D} as the input in this model. Qualitatively, the one without the damage field (blue line) can only learn the overall trend of a P-V curve both for training and testing cases, whereas the detailed pressure drop feature cannot be captured in the absence of the damage field in training. However, with the information from the damage field, the overall performance improves significantly, shown via the orange and red dash lines. As mentioned above, the true damage is a hidden quantity whereas only the predicted damage progression is available in practice. Hence, we plot the predicted P-V curve based on the true and predicted damage fields for further comparison. Although predictions based on the true data slightly outperform the predicted damage field, both capture the detailed underlying mechanics by matching better with the phase-field data, which implicitly indicates that a learned relationship exists between the structure and its corresponding mechanical property. 

\begin{table}
    \centering
    \begin{adjustbox}{width=0.7\textwidth}
    \begin{tabular}{c | c | c | c} 
    \hline
    Branch Net & Width & Trunk Net & Width \\ \hline
    \hline
    FNN & [50, 50, 50] & FNN & [50, 50, 50]\\
    \hline
    CNN & [4, 8, 12] & FNN & [50, 50, 50]\\
    \hline
    \end{tabular}
    \end{adjustbox}
    \caption{\textbf{Model architecture for the P-V curve.} The trunk net is a fully-connected neural network (FNN) with three hidden layers. The branch net can be a FNN or a convolutional neural network (CNN) with three blocks, where each block contains a convolutional layer, a LeakyReLU layer, and a batch normalization layer. [4, 8, 12] indicates the number of channels in each block.}
    \label{table:pv_network_size}
\end{table}

\begin{table}[]
\centering
\begin{adjustbox}{width=0.7\textwidth}
\begin{tabular}{c | c | c}
\hline\hline
\textbf{Training Errors} & $L_{2}^{2}$ Error & Relative Error \\
\hline \hline
DeepONet(Damage)  & $5.22\times10^{-4}$ & 3.10\%  \\
\hline
DeepONet(No Damage)  & $1.08\times10^{-3}$ & 4.15\%  \\
\hline
Linear Model  & $2.26\times10^{-3}$ & 6.48\%  \\
\hline \hline
\textbf{Testing Errors} & & \\
\hline \hline
DeepONet(True Damage)  & $7.55\times10^{-4}$ & 3.70\%  \\
\hline
DeepONet(Predicted Damage)  & $1.37\times10^{-3}$ & 4.88\%  \\
\hline
DeepONet(No Damage)  & $4.36\times10^{-3}$ & 8.62\%  \\
\hline
Linear Model & $2.47\times10^{-3}$ & 6.67\%  \\
\hline \hline 
\end{tabular}
\end{adjustbox}
\caption{Error Table for prediting P-V curves.}
\label{table:pv_error}
\end{table}

In Table~\ref{table:pv_error} we present all errors for the different models. In training, the $L_{2}^{2}$ error for DeepONet with the damage field in the input is $5.22\times10^{-4}$ and the relative error is approximately at 3\%. As baseline models, the DeepONet model without damage and a linear model are trained based on microstructures and P-V curves. Their $L_{2}^{2}$ error is significantly higher, on the order of $O(10^{-3})$, with relative errors are 4\% and 6\%, respectively. As for the testing errors, the DeepONet model using the true damage field has the lowest error at $7.55\times10^{-4}$ (3.7\%). In a more practical situation, the DeepONet model using predicted damage fields has a relatively low error at $1.37\times10^{-3}$ (4.88\%), exhibiting its practicality in a more realistic scenario. The baseline models, however, have the largest error with the relative error reach at 8.62\% and 6.67\%. More details of the linear model are included in the Appendix.

\section{Discussion} 
\label{sec:discussion}

In this paper, we demonstrate the potential of DeepONet for predicting the mechanical behavior of a heterogeneous aortic wall under dissection, here as reflected by the P-V curve and damage field in the case of a pressure-driven injection of fluid within the medial layer; we use phase-field finite element simulations as synthetic data. The model leverages recent advances in the operator learning model DeepONet. The predicted P-V curves agree well with the reference data produced by the phase-field finite element model. The damage field can be inferred based on the observable displacement field and its initial distribution of intralamellar radial struts, suggesting a potential approach to directly estimate the damage field from limited imaging data. Based on the inferred damage field, the predicted P-V curve has a smaller error than the baseline models. We also investigated the network structure and its impact on model generalization: a network with a smaller size tends to have a lower testing error. 

Practically, due to the extremely complicated mechanics, pure data-driven modeling without information on intralamellar damage progression would lead to an inaccurate inference of the P-V curve. A possible mitigation will be incorporating the underlying physics into the network, which transforms the framework to physics-informed DeepONet~\cite{wang2021learning, goswami2021physics}. In addition, our \textit{in silico} simulations only show dissection progression within the media layer, without developing in the radial direction. 

A real dissection, however, may yield three distinct outcomes: it may turn inward to form a false lumen or re-entry site, turn outward resulting in rupture, or stabilize, with possible subsequent healing or reinitiation. One possible reason that our model did not yield these different scenarios is our oversimplification of the real elastic lamellae structure: the lamellae are curly layers, not the straight layers (\ref{fig:multiphoton}). We are currently investigating such effects on the dissection progression and developing a new surrogate model for prediction. 
Another possible improvement would be to incorporate this model into a probabilistic framework, where uncertainty is modeled by predicting the mean and variance of the quantity of interest~\cite{wu2021parameter}. It is a natural framework in terms of describing a complicated bio-system with variability. Another promising extension of this work would be feature extraction and identification enabled by machine learning. From the phase-field simulations we can generate a correspondence between dissection progression and microstructure. It would then be possible to learn the relation between local structures and their contribution to the mechanical properties, which enables a more interpretable machine learning model.

\section*{Acknowledgment}
The work is supported by grant U01 HL142518 from the National Institutes of Health.

\section{Appendix}
\subsection{Prediction of Displacement Fields}

In this section, we present a network for predicting the displacement field based on the damage field at the current step. We aim to show the capability of the network to infer the displacement field given experimental measurements. The triplet of the training dataset is similar to that in Fig.~\ref{fig:damage_triplet} whereas the displacement and damage field are taken as the output and input, respectively. We use the same simulations as in section \ref{sec:Net_D} as training data.

\begin{figure}
    \centering
	\includegraphics[width=1.0\textwidth]{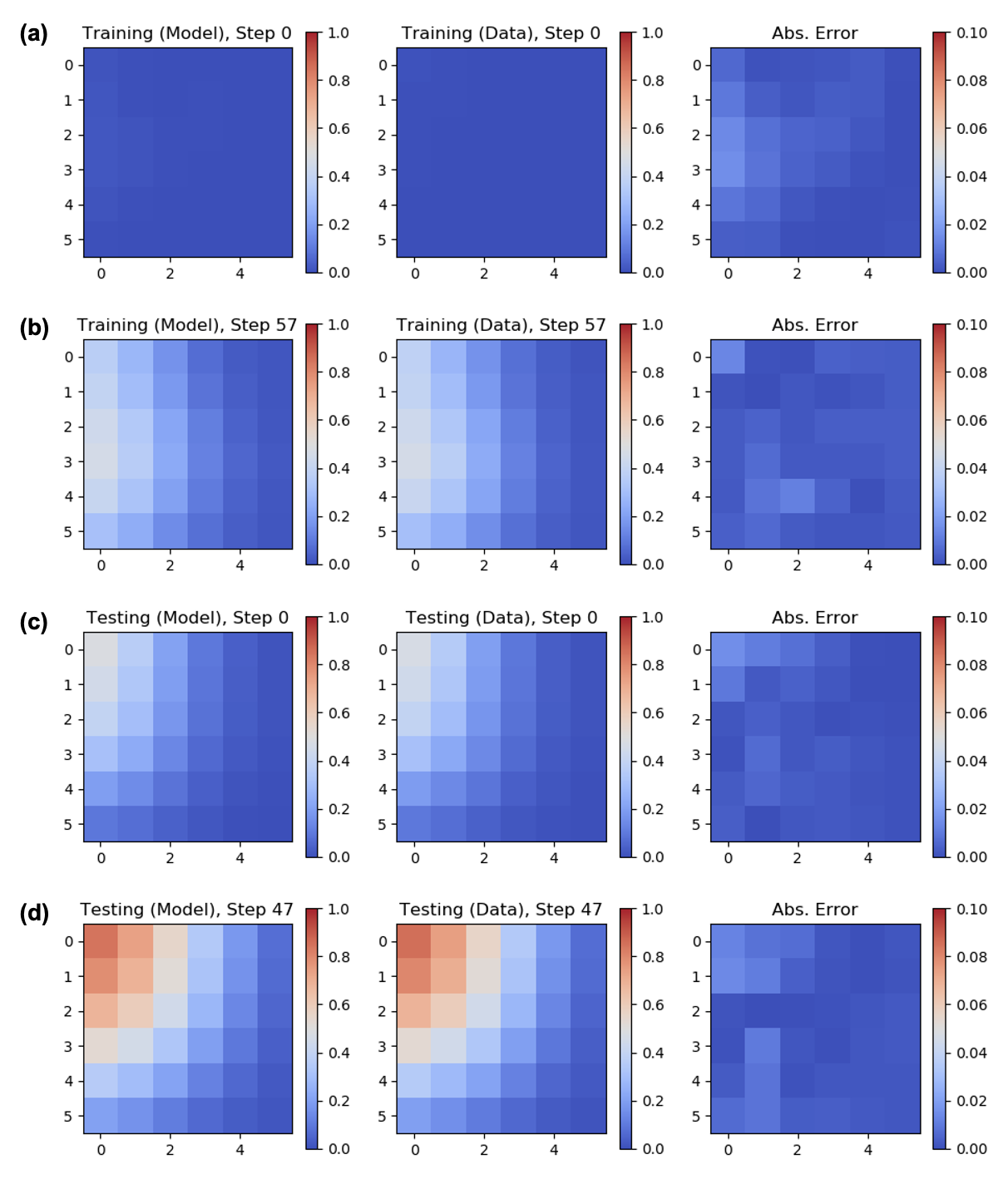} 
    \caption{\textbf{Training/Testing results of DeepONet for predicting displacement fields.} Model predictions and true displacement fields are shown in the first and second columns, while the third column presents their absolute difference. (a, b) show the results for a training case at step 0 and 57, while (c, d) represent for a testing case at step 0 and 47. The length unit is $\mu \textnormal{m}$.}
    \label{fig:displacement}
\end{figure}

We show the performance of DeepONet in predicting the displacement field in Fig.~\ref{fig:displacement}, where we plot the predictions in a similar fashion. The training data contains 1,900 cases with 200 additional cases for testing. We trained the model for 3,000 epochs on a single NVIDIA V100 GPU; (a) and (b) show the model prediction and true damage data at the initial and a later stage with the absolute error plotted in the third column. In (c, d), we compare the inference results for the displacement field from a testing case. The inferred displacement field matches the true data well, with the maximum pixel-wise relative error less than 5\%. 

\begin{figure}
    \centering
	\includegraphics[width=1.0\textwidth]{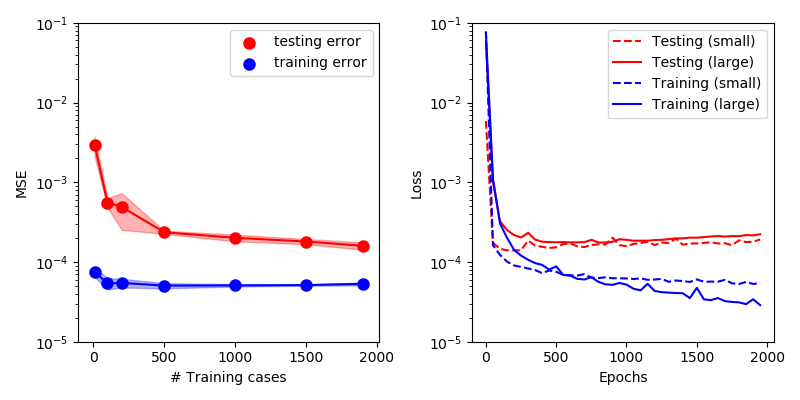} 
    \caption{\textbf{Mean square errors (MSE) for predicting displacement fields against the number of training cases and epochs.} The network architectures are the same in Table \ref{table:damage_network_size}. The error bars show one standard deviation from 5 runs with different initiations.}
    \label{fig:param_disp}
\end{figure}

Next, we investigate the number of training cases on the influence of prediction errors by training a network with a fixed architecture with differing amounts of data. With the increase of training data, the training error and testing error drop down to $2\times10^{-4}$ and $6\times10^{-5}$. The figure on the right shows the training/testing error against epochs for a large and small network. The large one has three hidden layers each with 100 neurons, whereas the small one contains three hidden layers and 50 neurons. By employing the small network, the testing error reduces from ~0.08 to 0.001 while the training cases increase from 10 to 1,900. The performance of the small network, indicated by the distance between training and testing error, is better than that of the large network.

\subsection{Field Predictions based on strut maps}

\begin{figure}
    \centering
	\includegraphics[width=0.7\textwidth]{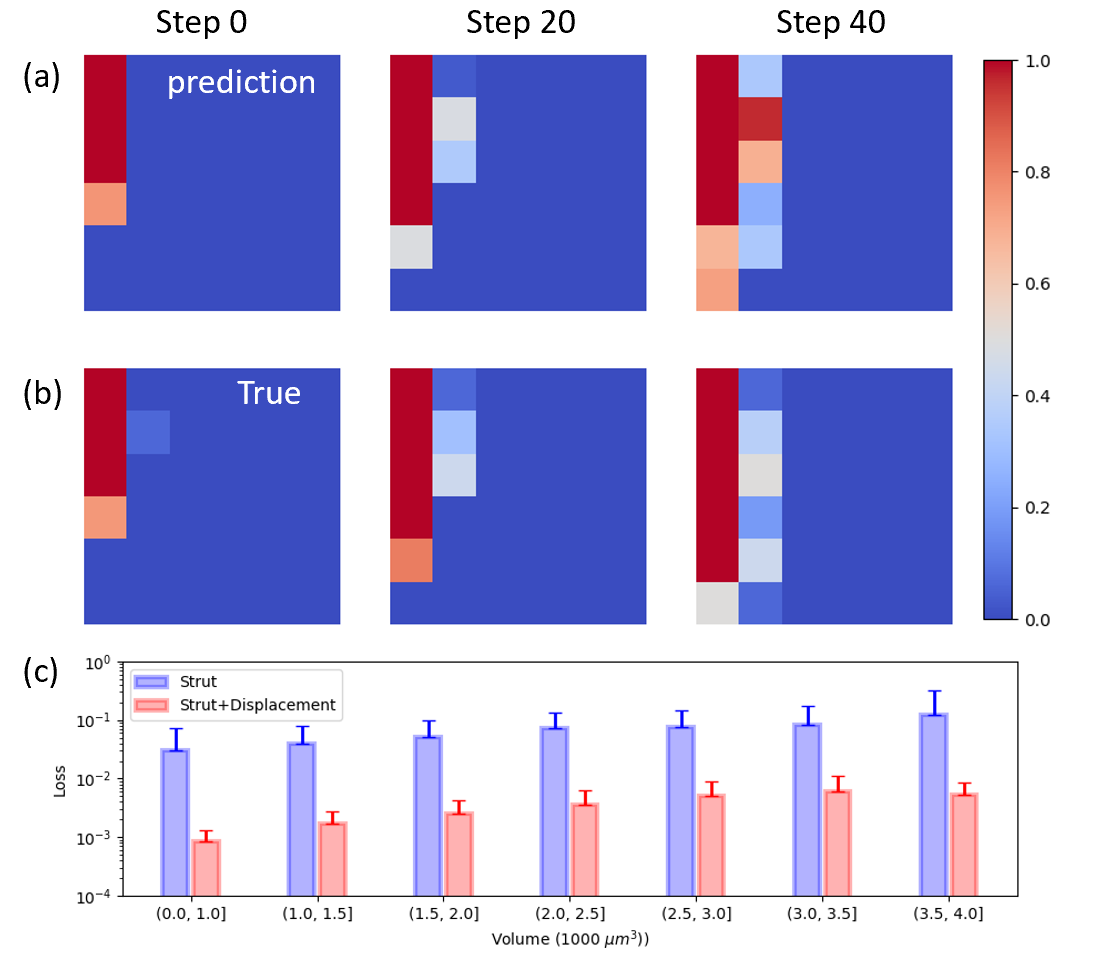} 
    \caption{\textbf{Testing results of DeepONet for predicting damage fields based on partial information.} (a) Model predictions and (b) true displacement fields at steps 0, 20, and 40 are shown, whereas the predictions are only based on strut maps. (c) Testing loss of models with different inputs versus injection volume. The red bars indicate testing loss ($L_{2}$ error) of model predictions with strut and damage field as input whereas the blue bars are predictions based only on strut maps. The error bars show the one standard deviation from 5 runs with different initializations.}
    \label{fig:app_dmg}
\end{figure}

\begin{figure}
    \centering
	\includegraphics[width=0.7\textwidth]{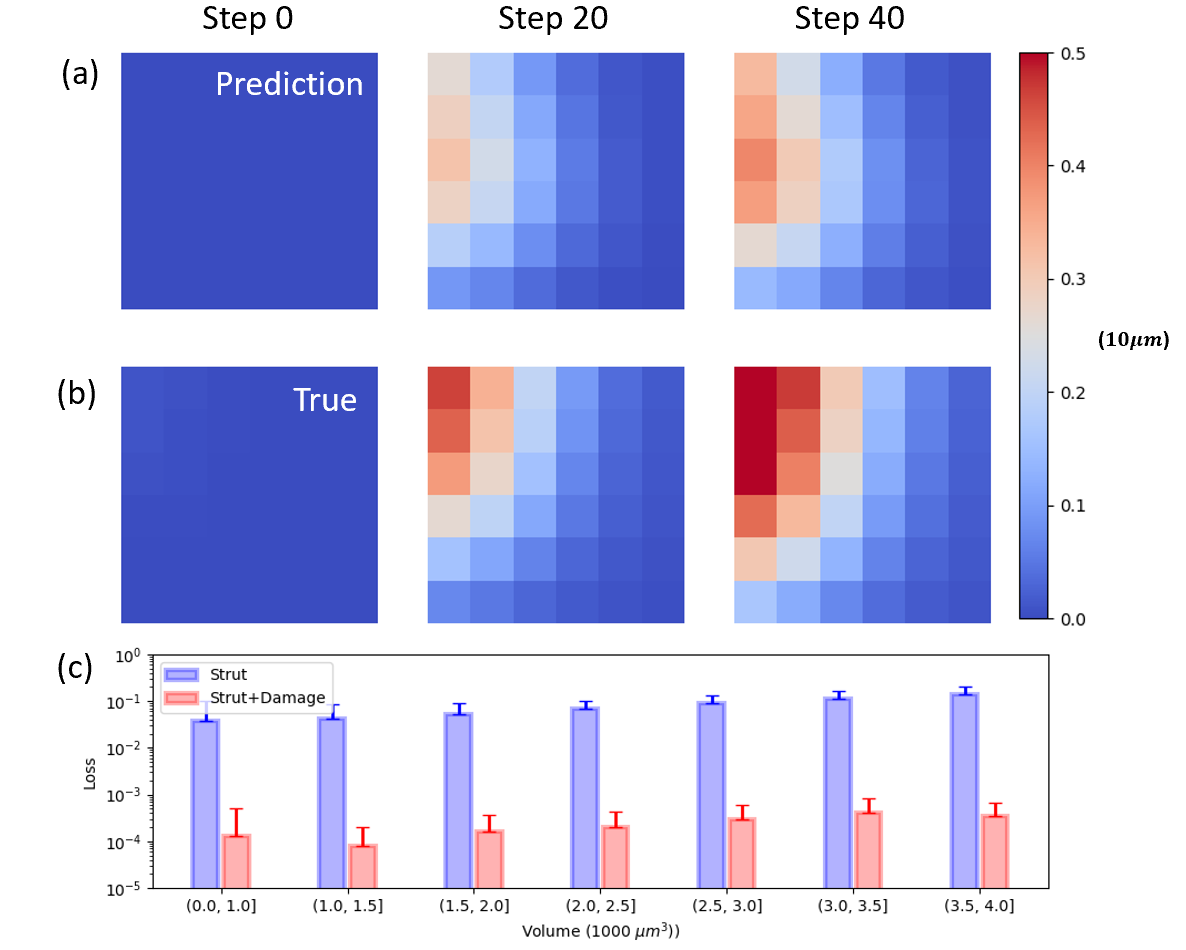} 
    \caption{\textbf{Testing results of DeepONet for predicting displacement fields based on partial information.} (a) Model predictions and (b) true damage fields at steps 0, 20, and 40 are shown, whereas the predictions are based only on strut maps. (c) Testing loss of models with different inputs versus injection volume. The red bars indicate testing loss ($L_{2}$ error) of model predictions with strut and damage fields as input whereas the blue bars are predictions only based on strut maps. The error bars show the one standard deviation from 5 runs with different initializations.}
    \label{fig:app_disp}
\end{figure}

The ultimate goal is to develop a reliable surrogate model of dissection progression only with the information of strut maps for each subject. However, due to the nature of this complex system, data-driven modeling seldom renders a satisfactory and reliable result without the guidance of additional information, e.g., the damage or displacement field. In this section, we conduct a quantitative comparison between models with partial (strut) and additional (displacement/damage) information. In Fig.~\ref{fig:app_dmg}, we show the testing results for predicting damage fields based on strut maps. Fig.~\ref{fig:app_dmg} (a) shows the current damage field at steps 0, 20, and 40 from a model with strut maps. These results can qualitatively match the true damage progression at an early injection stage shown by a small injection volume, but the error will drastically grow at later stages (step 40 or later). A more quantitative result is presented in (c) where testing errors for the two models (with an error bar) show the overall performance over different injection stages. In general, both the blue bars, representing predictions based on the strut map, and the red bars show a growth from a small to a large injection volume. However, with the guidance of additional information, i.e., displacement fields, the network is able to achieves a much better prediction with orders-of-magnitude lowering of testing errors. A similar trend is observed in Fig.~\ref{fig:app_disp} for predicting displacement fields. 

\subsection{Exploratory Analysis of Pressure--Volume Data}

Because the data-generating finite element simulations of dissection were executed until failure, the training and testing data used to develop and validate the \textit{Net P-V} presented herein covered disparate ranges of injection volume and pressure values. To examine the statistical characteristics of the resulting training and testing data, we quantified the empirical distributions of the observed volume and pressure values, as well as several moments of those distributions. Hence, we generate another dataset with 1,000 cases using the same approach as that for the main text. However, the size of strut maps are now $36\times 32$.

\begin{figure}
    \centering
	\includegraphics[width=1.0\textwidth]{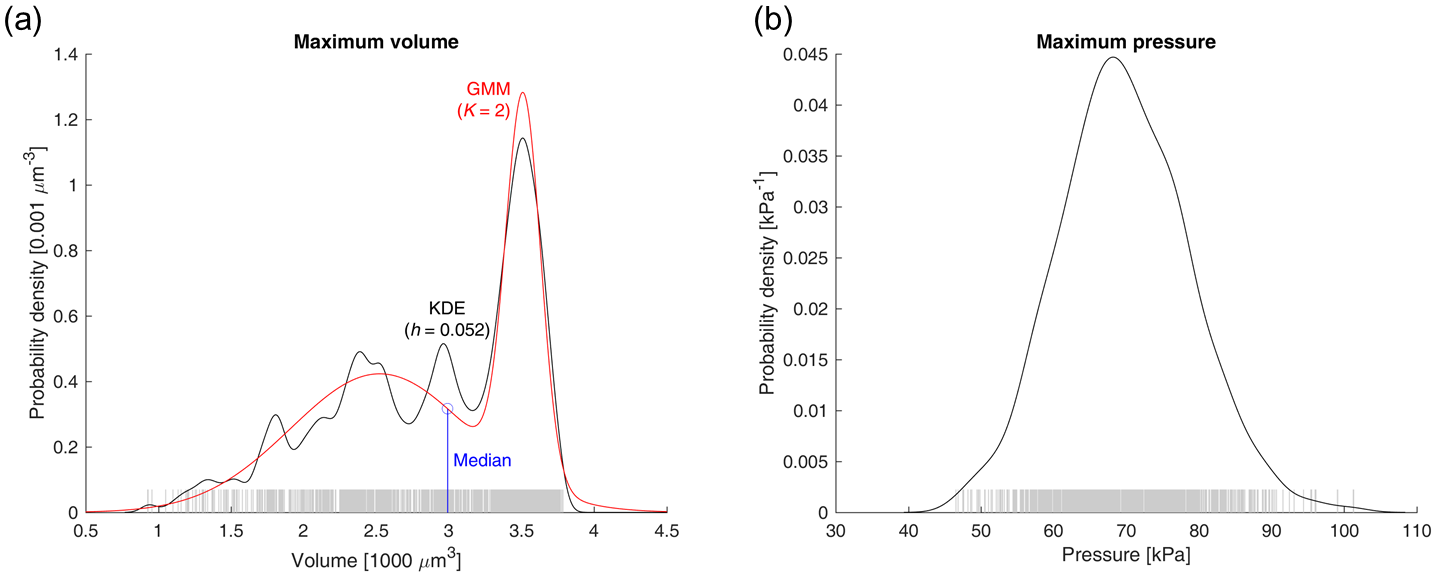} 
    \caption{
    \textbf{Maximum volume and pressure distributions.} Because the data-generating finite element simulations of dissection were executed until failure, the training and test data used to develop and validate the P-V predictive models presented herein exhibited a large range of maximum injection volume and pressure values. (a) Kernel density estimation of the maximum volume data yielded two primary sub-populations of samples: one tightly concentrated around $\sim$3500~$\mu \textnormal{m}^3$ and another more variable group of samples that failed earlier. The overall distribution is well described by a two-component Gaussian mixture. (b) The distribution of maximum pressure values was unimodal and nearly Gaussian. GMM, Gaussian mixture model; KDE, kernel density estimation}
    \label{fig:max_volume_pressure_distributions}
\end{figure}

Diffusion-based kernel density estimation~\cite{botev2010kernel} (optimal bandwidth $h = 52~\mu \textnormal{m}^3$) of the maximum injection volume attained by the finite element simulations yielded two primary sub-populations of samples: one tightly concentrated around $\sim$3500~$\mu \textnormal{m}^3$ and another more variable group of samples that mostly experienced failure at lower injection volumes (Fig.~\ref{fig:max_volume_pressure_distributions}(a)). We found that the overall distribution of maximum volume values was well described by a Gaussian mixture model with $K=2$ components. In contrast, the distribution of maximum pressure values was unimodal and nearly Gaussian (Fig.~\ref{fig:max_volume_pressure_distributions}(b)) with mean value at ~68 kPa.

\begin{figure}
    \centering
	\includegraphics[width=0.9\textwidth]{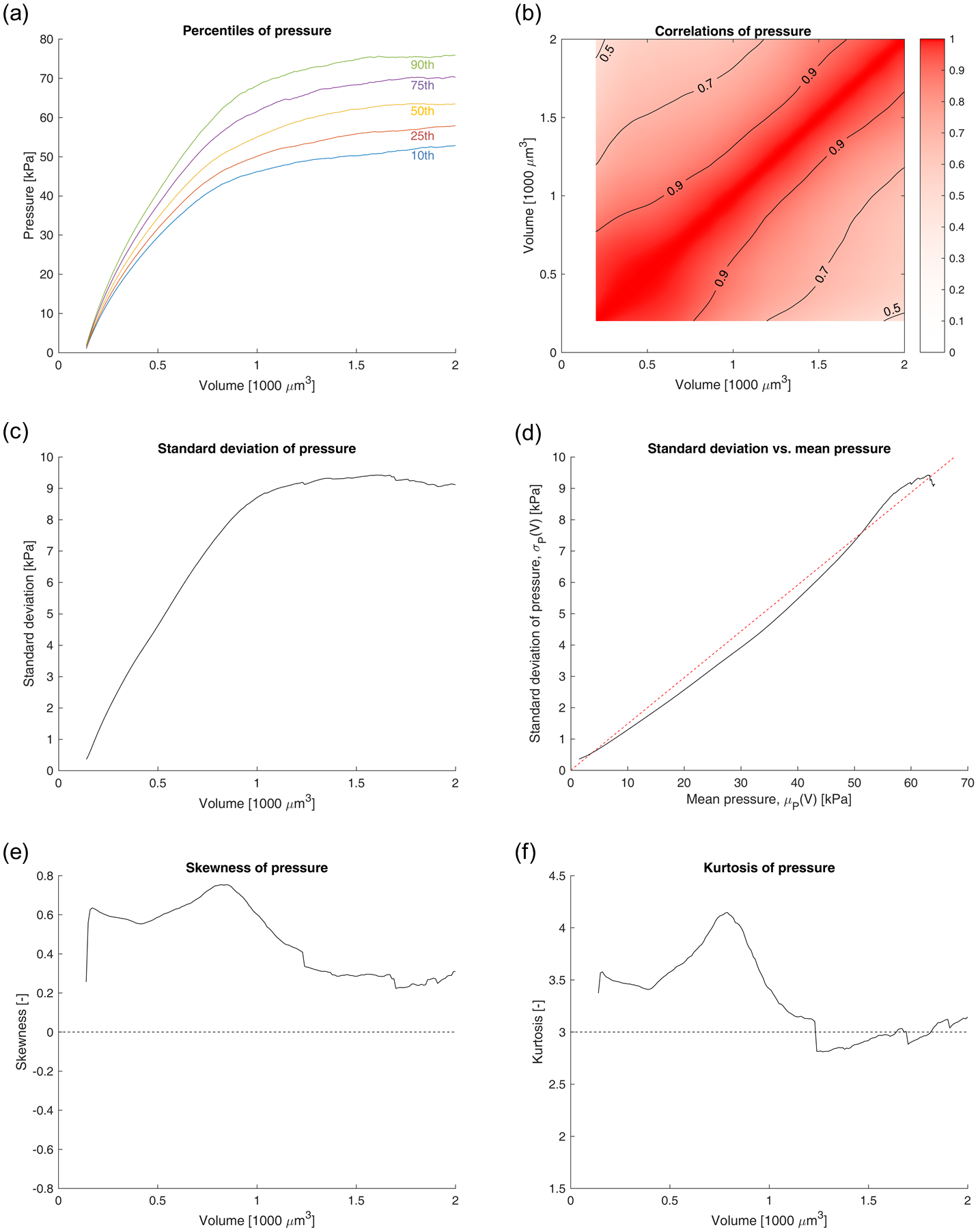} 
    \caption{
    \textbf{Statistical analysis of the pressure-volume data.} (a) Percentiles and (b) correlations of observed pressure as a function of the injection volume. (c) Standard deviation of observed pressure as a function of the injection volume. (d) Standard deviation of volume-specific pressure plotted against the mean pressure observed at the same volume, showing a nearly linear relation. (e) While at low volumes, the pressure data is positively skewed, the skewness is closer to zero at higher volumes. (f) Similarly, pressure distributions are leptokurtic at low volumes, but approach a kurtosis of 3 at higher volumes. Together, these metrics suggest that the pressure distributions at high volumes are nearly Gaussian (cf. Fig.~\ref{fig:max_volume_pressure_distributions}(b)).}
    \label{fig:pressure_data_statistics}
\end{figure}

To further quantify the distribution of observed injection pressures, we examined various percentiles and correlations of the pressure data as a function of the injection volume (Fig.~\ref{fig:pressure_data_statistics}(a,b)). We found that the effective width of the volume-specific pressure distribution increased monotonically with injection volume, which can be summarized through a corresponding monotonic increase in the standard deviation (Fig.~\ref{fig:pressure_data_statistics}(c)). Moreover, we found that the standard deviation of the pressure distribution was linearly proportional to the corresponding mean observed pressure at the same injection volume (Fig.~\ref{fig:pressure_data_statistics}(d)). The pressure data were positively skewed especially at low injection volumes, though more symmetric as volume (and mean pressure) increased (Fig.~\ref{fig:pressure_data_statistics}(e)). Similarly, pressure distributions were leptokurtic at low volumes, but approached a kurtosis of 3 at higher volumes. Together, these metrics suggest that the pressure distributions at high volumes are nearly Gaussian (Fig.~\ref{fig:pressure_data_statistics}(f)), consistent with our qualitative examination of the maximum pressure distribution (Fig.~\ref{fig:max_volume_pressure_distributions}(b)). 

\subsection{Linear Model Prediction of the Pressure--Volume Curve}
\begin{figure}
    \centering
	\includegraphics[width=1.0\textwidth]{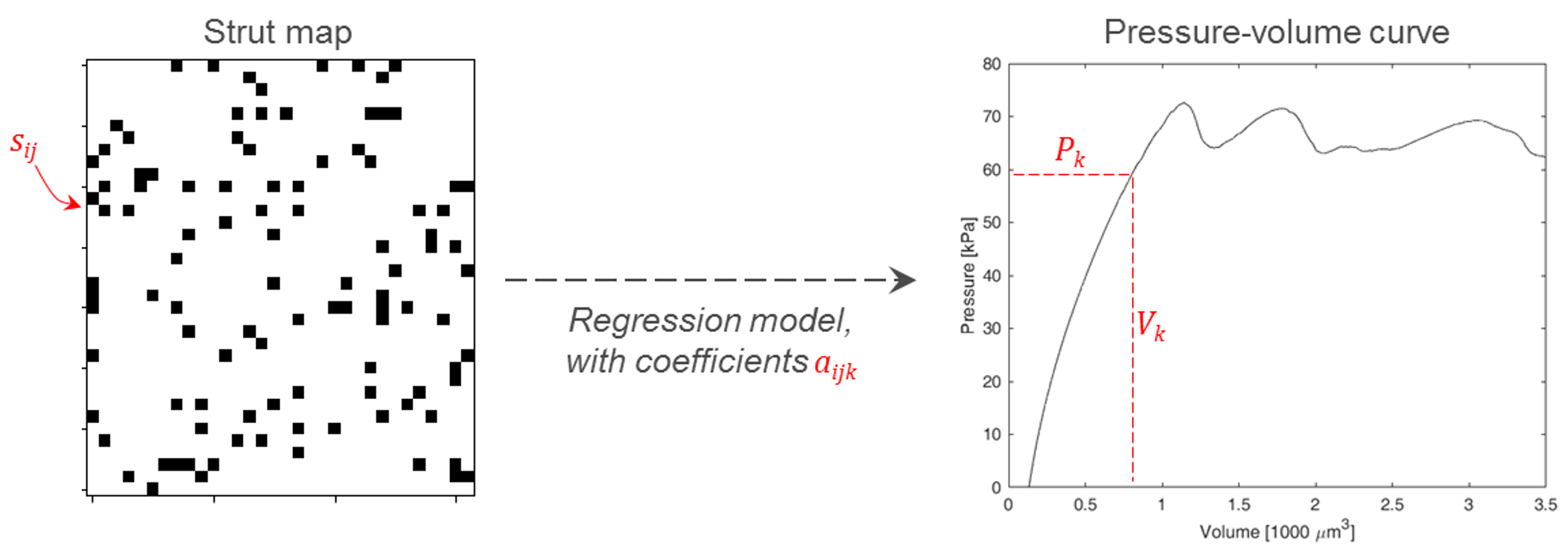} 
    \caption{
    \textbf{Summary of the linear P-V model.} At each volume $V_k$, the corresponding pressure $P_k$ is predicted from the elastin strut map $\mathbf{s}$ and a set of volume-specific linear regression coefficients $\mathbf{a}_k$. Elastin struts ($s_{ij} = 1$) are displayed in black, while non-elastin matrix ($s_{ij} = 0$) pixels are displayed in white.}
    \label{fig:linreg_approach}
\end{figure}

Herein, we present a linear regression model to predict P-V data from the elastin strut map, which can serve as a (low-fidelity) baseline model for comparison with our higher-fidelity DeepONet results. In this approach, the inner product between the array of tissue microstructure $\mathbf{s}$, where $s_{ij} = 0$ to indicate non-elastin matrix and $s_{ij} = 1$ to indicate the presence of an elastin strut, and a coefficient array $\mathbf{a}$ was used to predict the observed pressure $P$ as a function of injection volume $V$. Specifically, at a particular injection volume $V_k$,
\begin{equation}
\begin{aligned}
    P_k - \bar{P}_k & = \sum_i \sum_j a_{ijk} \left( s_{ij} - \bar{s}_{ij} \right) + \varepsilon_k \\
    \textnormal{for } \bar{P}_k & = \mathbb{E} \left[ P_k \right] \approx \frac{1}{N} \sum_{n=1}^N \left( P_k \right)_n \\
    \textnormal{and } \bar{s}_{ij} & = \mathbb{E} \left[ s_{ij} \right] \approx \frac{1}{N} \sum_{n=1}^N \left( s_{ij} \right)_n,
\end{aligned}
\end{equation}
where $\mathbb{E}[\,\cdot\,]$ denotes the expected value; $\bar{P}_k$ and $\bar{s}_{ij}$ are the expected values of the pressure and the tissue microstructure indicator function respectively, each estimated from the sample means of the training data; $n$ is the training sample index; $N$ is the training data sample size; and $\varepsilon_k \sim \mathcal{N}(0, \, \sigma_{k}^2)$ is the pressure residual unexplained by a linear function of the strut map input (Fig.~\ref{fig:linreg_approach}). The model coefficients $a_{ijk}$ were determined directly via least-squares linear regressions of the volume-specific P-V data.

\begin{figure}
    \centering
	\includegraphics[width=1.0\textwidth]{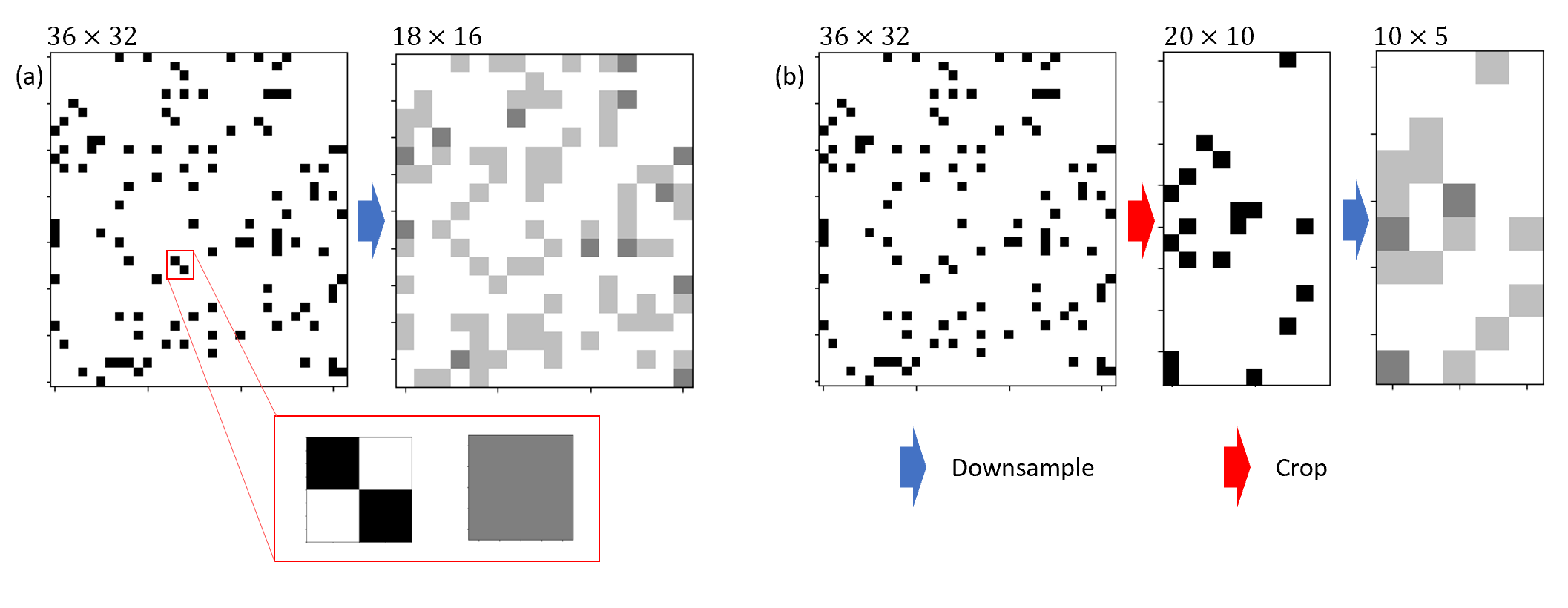} 
    \caption{
    \textbf{Preprocessing of the strut map for linear regression.} (a) In the first version of the model, the input elastin strut map data is downsampled to half of its size along each dimension. In downsampling, each adjacent 2~$\times$~2 block of pixels in the strut map is reduced to one pixel, whose value is set to the mean in the original pixel block. (b) In the second version of the model, the original strut map is first cropped to include only the top-left region surrounding the injection, and then the cropped strut map is downsampled, leading to a substantially reduced input space.}
    \label{fig:linreg_downsampling}
\end{figure}

To examine the effects of varying complexity in the input space, we applied the above linear model to two preprocessed versions of the elastin strut map: (1) a version of the strut map that was downsampled to half its size along each dimension (Fig. \ref{fig:linreg_downsampling}(a)), and (2) a version of the strut map that was first cropped to only the 10~$\times$~5 top-left region surrounding the initial injection site and then downsampled as in the first version (Fig. \ref{fig:linreg_downsampling}(b)). In downsampling, each adjacent 2~$\times$~2 block of pixels in the strut map is reduced to one pixel, whose value is set to the average fraction of elastin in the original pixel block (this is equivalent to the application of a box filter to the original strut map). The total input size was thus 288 in the first version of the model but only 50 in the second version. The additional cropping step served to substantially reduce the input space, thus limiting the complexity of the model and reducing the potential for the model to overfit the training data while performing more poorly on the testing data. Alternatively, a regularization scheme could have been incorporated (e.g. via ridge or lasso regression); however, such approaches typically penalize coefficient magnitudes even in regions where the optimal coefficients are ``legitimately'' (i.e. non-spuriously) high. In contrast, cropping leaves those coefficients unpenalized while still removing spurious coefficients from distant tissue regions that should not be expected to contribute to the delamination response.

\begin{figure}
    \centering
	\includegraphics[width=1.0\textwidth]{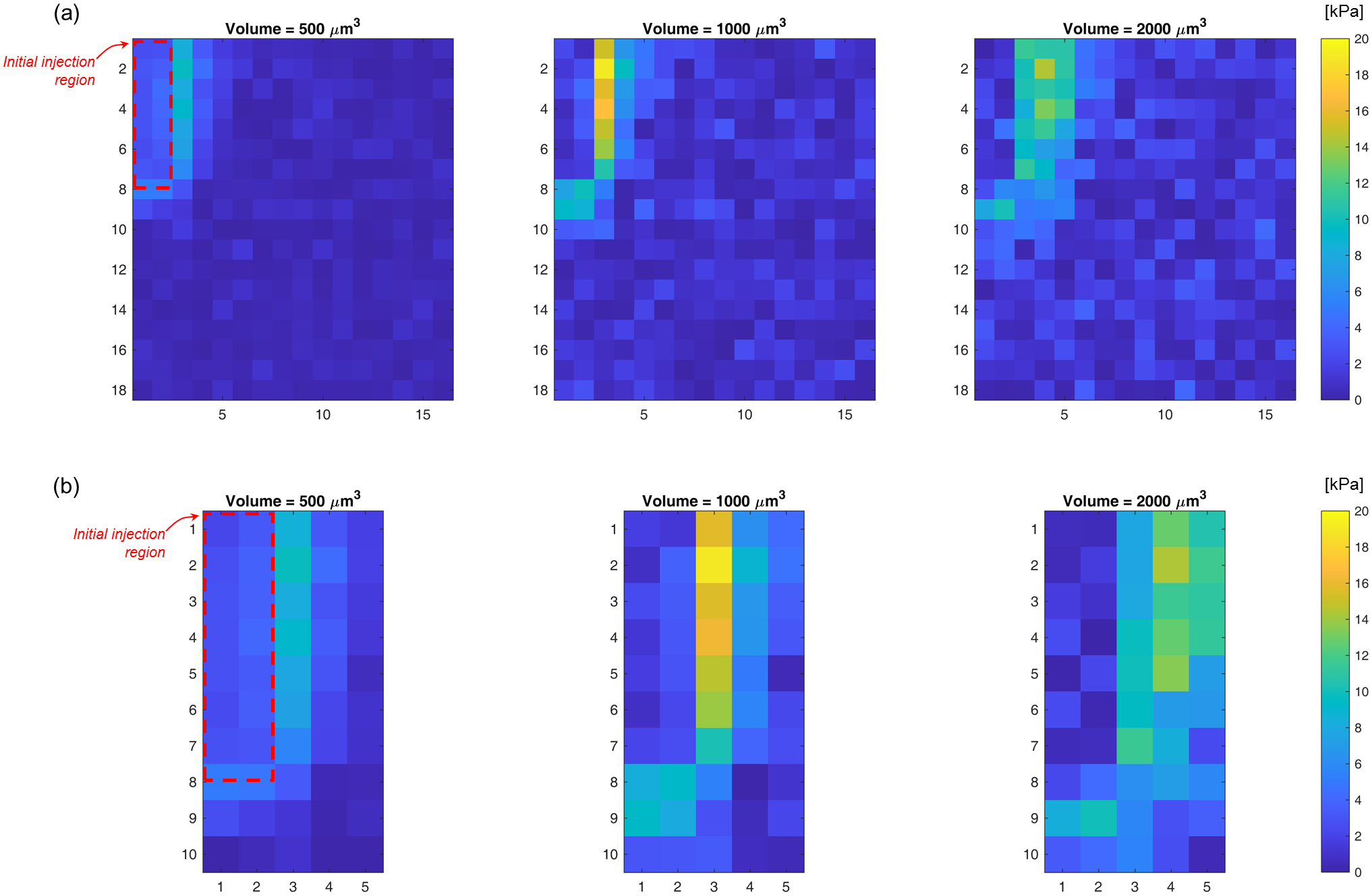} 
    \caption{
    \textbf{Absolute values of the least-squares linear P-V model coefficients.} (a) Due to the large input space in the first version of the model, the regression results in spurious coefficient values in regions far from the injection site, suggesting possible overfitting. (b) The reduced input space yielded similar results in the remaining coefficients. In both models, pressure predictions at low injection volumes (e.g. $\leq 1000~\mu \textnormal{m}^3$) are dominated by the elastin microstructure along the immediate boundary of the initial injection region, but broader regions of the tissue contribute to the model at higher volumes ($\sim$2000$~\mu \textnormal{m}^3$).}
    \label{fig:linreg_coefficients}
\end{figure}

Regression results from both versions of the linear model yielded consistent trends in the corresponding optimal coefficient maps (Fig.~\ref{fig:linreg_coefficients}). In both models, pressure predictions at low injection volumes (e.g. $\leq 1000~\mu \textnormal{m}^3$) were dominated by the elastin microstructure along the immediate boundary of the initial injection region. At higher volumes ($\sim$2000~$\mu \textnormal{m}^3$), after the injection region has enlarged in a sample-specific manner, broader regions of the tissue contribute to the model prediction, though the dominant coefficients still neighbor the (current) injection region. Due to the large input space in the first version of the model (downsampled only), the regression also results in spurious coefficient values in regions far from the injection site, suggesting possible overfitting of the training data.

\begin{figure}
    \centering
	\includegraphics[width=1.0\textwidth]{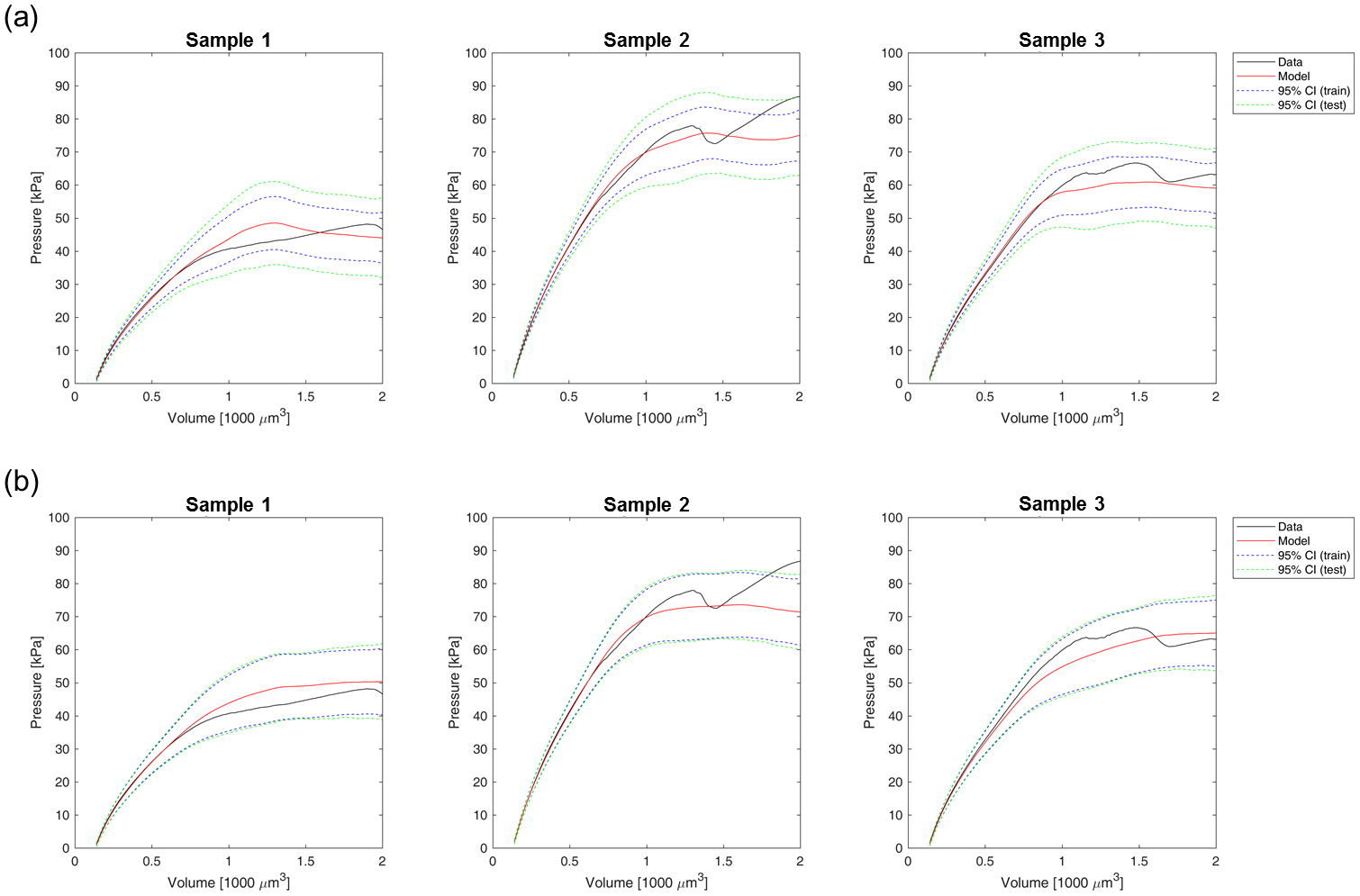} 
    \caption{
    \textbf{Performance of the linear P-V model for three representative samples.} (a) In the first version of the model, the uncertainties in predicted pressure values estimated from the training data are substantially smaller than the errors observed in validation against the test data, indicating that the model has overfit the training data. (b) In the second version of the model, which utilizes a cropped and downsampled strut map as its input, the uncertainties estimated from the training and test data are approximately equal. CI, credible interval}
    \label{fig:linreg_overfitting}
\end{figure}

To assess the performance and robustness of each version of the linear model, we examined the discrepancy in the distribution of model residuals for both the training and testing data. In the first version of the model, where the input strut map data was only downsampled, the residuals in pressure prediction were substantially larger in the testing data compared to the training data (Fig.~\ref{fig:linreg_overfitting}(a)), indicating that the model overfit the training data. In contrast, training and testing residuals were almost identical in the second version of the model, where the input strut map data was both cropped and downsampled (Fig.~\ref{fig:linreg_overfitting}(b)). This improved performance was especially pronounced at higher injection volumes, where the average ratio between the root-mean-square testing and training residuals was $\sim$1.6 in the downsampled-only model, but only $\sim$1.1 in the downsampled and cropped model.

\begin{figure}
    \centering
	\includegraphics[width=1.0\textwidth]{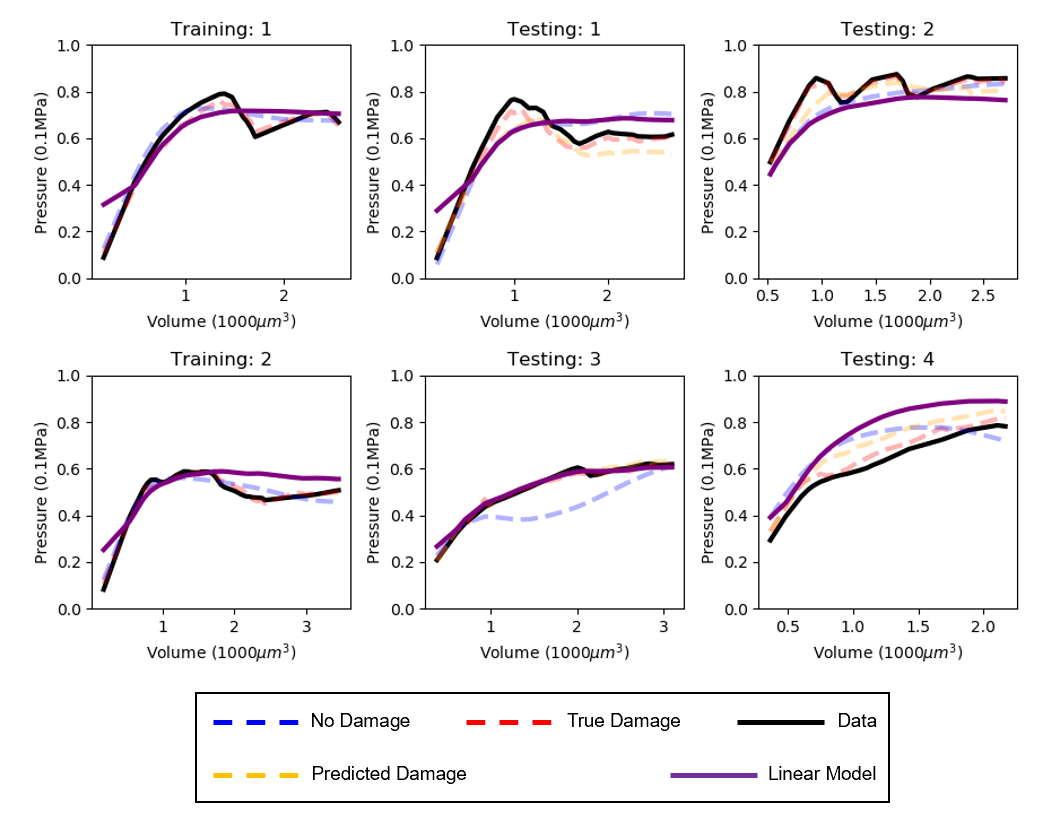}
    \caption{\textbf{DeepONet for predicting P-V curves.} Model predictions (dash lines) for training cases 1 and 2 are compared with the true data (black line, first column). In particular, model predictions with and without damage fields are plotted in red and blue, respectively. The other columns compare the testing data with models trained with (red)/without (blue) the current damage field. The branch net is FNN trained with 1900 training cases. The true data were generated from the phase-field finite element solver.}
    \label{fig:pv_curve_app}
\end{figure}

Fig.~\ref{fig:pv_curve_app} compares predictions of the linear model and the DeepONet model proposed in the main text. The black solid lines represent the P-V curves from the finite-element simulations while the purple line denotes the mean linear model prediction. As mentioned, the predictions from the linear model can only capture the overall trend of the P-V curve, failing to characterize the detailed pressure drop induced by struts that break. Results from other models are also plotted as a comparison. A more quantitative comparison of errors for the linear model is in Table~\ref{table:pv_error}.

\subsection{Iterative Prediction of the Damage Field Using Logistic Regression}
\label{sec:appendix:logistic_regression}
Herein, we present a logistic regression model to iteratively predict the damage field data from the elastin strut map and predicted damage values at preceding injection volumes. This model was developed to serve as a baseline/proof-of-concept model for comparison with our higher-fidelity DeepONet results. In this approach, given a design matrix $\mathbf{X}$ describing the current tissue microstructure and a set of model parameters $\mathbf{\beta}$, the probability of a particular pixel being the next to experience damage is modeling by a logistic function using
\begin{equation}
    \mathrm{Pr} \left( \mathrm{tear} \right) = \frac{1}{1 + e^{-\mathbf{X \beta}}}.
\end{equation}

\begin{figure}
    \centering
	\includegraphics[width=1.0\textwidth]{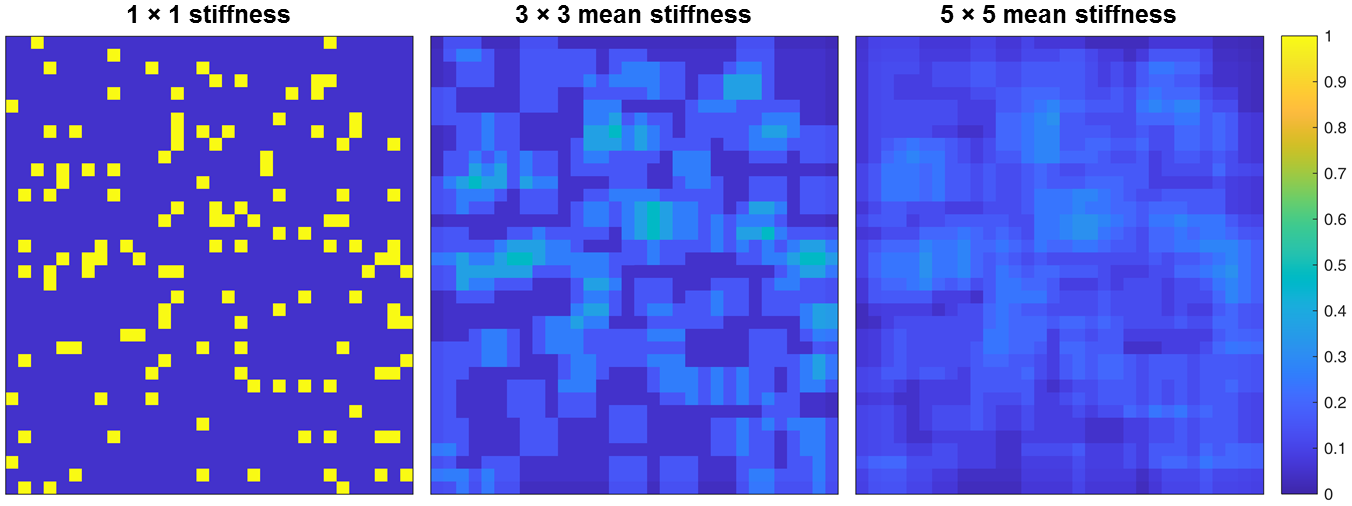}
    \caption{\textbf{Neighborhood normalized stiffness maps.} Maps of the tissue stiffness averaged over 1$\times$1, 3$\times$3, and 5$\times$5 pixel neighborhoods, for a representative sample. The stiffness values were computed directly from the elastin strut map, and normalized to the prescribed elastin stiffness, which is 20$\times$ greater than that of the non-elastin matrix.}
    \label{fig:logreg_stiffness}
\end{figure}

From our mechanistic understanding of the underlying tear propagation, we included in $\mathbf{X}$ parameters related to the current damage state of the tissue surrounding the pixel of interest as well as a quantification of the local tissue stiffness, both of which are known to influence the instantaneous tearing location and direction \cite{ban2021differential}. The local damage state was described using one-hot encodings of the number of lateral and diagonal neighbors that were currently damaged, while the local stiffness was described using 1$\times$1, 3$\times$3, and 5$\times$5 neighborhood averages of the underlying tissue constituent stiffness moduli, to account for the distance-dependent effect of local stiffness. Including a constant term, this results in 10 model parameters $\beta_i$ corresponding to the design matrix
\begin{equation}
    \mathbf{X} = 
    \begin{bmatrix}
        1 \\ \mathbb{1}_1 \left( n_\text{dam,\,lat} \right) \\ \mathbb{1}_2 \left( n_\text{dam,\,lat} \right) \\ \mathbb{1}_3 \left( n_\text{dam,\,lat} \right) \\ \mathbb{1}_1 \left( n_\text{dam,\,diag} \right) \\ \mathbb{1}_2 \left( n_\text{dam,\,diag} \right) \\ \mathbb{1}_3 \left( n_\text{dam,\,diag} \right) \\ \mu_{1 \times 1} \\ \mu_{3 \times 3} \\ \mu_{5 \times 5}
    \end{bmatrix}^\mathrm{T},
\end{equation}
where $\mathbb{1}_a(x)$ is the indicator function (equal to 1 if $x=a$ and 0 otherwise), $\{n_\text{dam,\,lat}, \, n_\text{dam,\,diag}\}$ are the number of damaged lateral and diagonal neighbors, and $\{\mu_{1 \times 1}, \, \mu_{3 \times 3}, \, \mu_{5 \times 5}\}$ are the local tissue stiffness values averaged over the corresponding neighborhood sizes. The stiffness values were normalized to the stiffness of elastin and the stiffness of the non-elastin matrix was prescribed as $1/20^\text{th}$ of the elastin stiffness following \cite{ban2021differential} (Fig.~\ref{fig:logreg_stiffness}).

\begin{figure}
    \centering
	\includegraphics[width=1.0\textwidth]{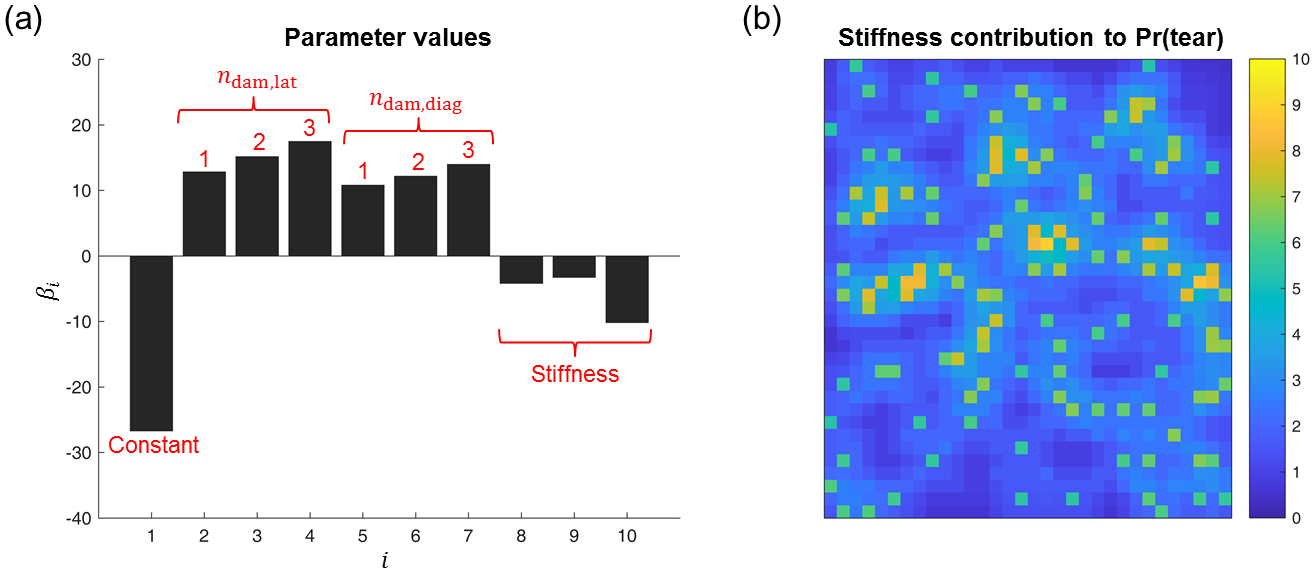}
    \caption{\textbf{Logistic regression results.} (a) Best-fit coefficients for the logistic regression model, showing contributions of damaged neighboring pixels and effective local tissue stiffness. (b) Corresponding map of the contribution of local tissue stiffness to the probability of tearing.}
    \label{fig:logreg_results}
\end{figure}

The resulting best-fit coefficients showed that the number of currently damaged neighbors and the local tissue stiffness both contribute substantially to the probability that a given pixel will be the next to experience damage---that is, to tear (Fig.~\ref{fig:logreg_results}(a)). An increase in the number of currently damaged neighbors, whether lateral or diagonal, was associated with an increased probability of tearing pressure $\mathrm{Pr}(\mathrm{tear})$. In contrast, greater local tissue stiffness was associated with a smaller $\mathrm{Pr}(\mathrm{tear})$. Note that while the absolute value of $\beta_{10}$ is largest among the stiffness-associated parameters, this does not imply that the average $5\times5$ stiffness is most predictive of tearing, since the contribution of this parameter is spread over 25 total pixels. Rather, the relative contribution of each neighborhood to $\mathrm{Pr}(\mathrm{tear})$ is computed by the product of the corresponding parameter and the neighborhood size. The resulting map of the effective contribution of stiffness to the probability of tearing is heterogeneous but fairly diffuse (Fig.~\ref{fig:logreg_results}(b)), highlighting the length scale of which the differential stiffness acts. For an elastin strut surrounded entirely by non-elastin matrix, the normalized stiffness contribution to decreasing $\mathrm{Pr}(\mathrm{tear})$ from the elastin pixel itself is 4.944, while the remaining $3\times3$ neighborhood pixels contribute 0.767 each and the still-remaining $5\times5$ neighborhood pixels contribute 0.406 each, consistent with the stiffness effect decreasing monotonically with distance.

\begin{figure}
    \centering
	\includegraphics[width=0.8\textwidth]{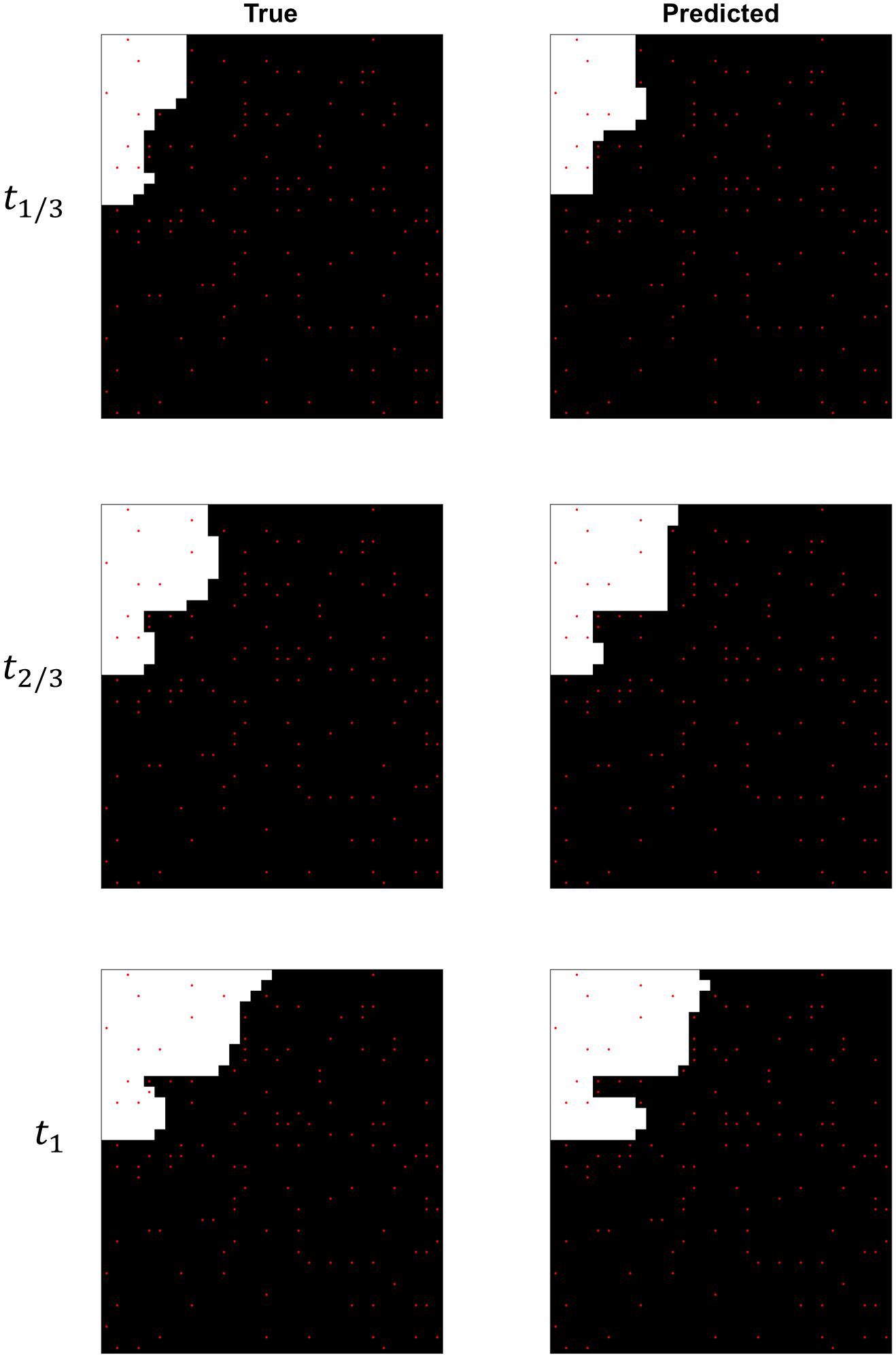}
    \caption{\textbf{Logistic regression-based damage prediction.} True (i.e. finite element-derived) and predicted (by the logistic regression model) maps of damage, shown at frames one-third ($t_{1/3}$) and two-thirds ($t_{2/3}$) of the way through the simulation as well as at the final simulation time ($t_1$). The damaged region is displayed in white, while the locations of elastin struts are shown as red dots.}
    \label{fig:logreg_prediction}
\end{figure}

To evaluate the predictive capability of the model, we numerically simulated the iterative propagation of damage, starting from the initial injection region. In each time step, $\mathrm{Pr}(\mathrm{tear})$ was computed for every yet-undamaged pixel, using the elastin strut map and the \textit{predicted} damage field from the previous time step as predictors, then the damage was assigned to the pixel with the highest $\mathrm{Pr}(\mathrm{tear})$. While the order in which pixels experienced damage often differed from the ``ground-truth'' finite element results, the evolution of the predicted damaged region qualitatively matched the shape of the true damaged region (Fig.~\ref{fig:logreg_prediction}). Note especially that this model, while simple in its construction, predicted correctly that damage is substantially less likely---and thus delayed---in areas where there is a high density of elastin struts, particularly when those struts are spatially aligned and form a ``barrier'' to tear propagation.

On the whole, the results from this logistic regression-based predictive model suggest that the damage field can be well described by accounting primarily for the state of the tissue microstructure along the immediate boundary of the current damaged region, though a more complex modeling framework is needed to improve accuracy in predicting the nuanced patterns in the evolving tear propagation and the corresponding pressure--volume relation. In the DeepONet approach, we thus prescribed a convolutional neural network (CNN) architecture to the trunk net, since this architecture similarly relies disproportionately on interactive effects between small neighborhoods of pixels (section~\ref{sec:pv_curve}).

\bibliographystyle{abme}
\bibliography{reference}
\end{document}